\documentclass{aa}
\usepackage{color}
\usepackage{etex}
\usepackage{graphicx}
\usepackage{lscape}
\usepackage{morefloats}
\usepackage{natbib}
\usepackage[scaled]{helvet}
\usepackage[varg]{txfonts}
\usepackage{url}
\usepackage{xspace}
\bibpunct{(}{)}{;}{a}{}{,}
\newcounter{Rco}
\newcommand{\ionw}[3]{\mbox{\ion{#1}{#2}~$\lambda\,#3\,\mathrm{\AA}$}\xspace}

\newcommand{\Jonw}[3]{\mbox{\ion{#1}{#2}~$\lambda\,#3$\,\AA}\xspace}

\newcommand{\logg}{\mbox{$\log g$}\xspace}
\newcommand{\loggw}[1]{\mbox{$\log g\hspace{-0.5mm} =\hspace{-0.5mm}  #1$}}

\newcommand{\Teff}{\mbox{$T_\mathrm{eff}$}\xspace}
\newcommand{\Teffw}[1]{\mbox{$\Teff\hspace{-0.5mm} =\hspace{-0.5mm} #1 \,\mathrm{K}$}}
\newcommand{\ebv}{\mbox{$E_{B-V}$}}

\newcommand{\Msol}{$M_\sun$\xspace}

\newcommand{\aador}{\object{AA\,Dor}\xspace}
\newcommand{\cp}{\object{CPD$-69\degr 389$}\xspace}
\newcommand{\hd}{\object{HD\,269696}\xspace}
\newcommand{\lb}{\object{LB\,3459}\xspace}
\newcommand{\hw}{\object{HW\,Vir}\xspace}
\begin{document}

\title{Search with UVES and XSHOOTER 
       for signatures of the low-mass secondary
       in the post common-envelope binary \object{AA\,Dor}
       \thanks{Based on observations collected at the European Southern Observatory, Chile, 
               programs 066.D-1800 and 092.C-0692.
              }$^,$          
       \thanks{Based on observations made with the NASA-CNES-CSA Far Ultraviolet Spectroscopic Explorer.
              }$^,$
       \thanks{Figures 
               \ref{fig:uves} - \ref{fig:nir}, 
               \ref{fig:detlimit}, and \ref{fig:tlisainput} - \ref{fig:nir_17590-17710}, 
               are only available in electronic form via http://www.edpsciences.org.
              }
      }

\titlerunning{Search for \object{AA\,Dor\,B}}

\author{D\@. Hoyer\inst{1}
        \and
        T\@. Rauch\inst{1}
        \and
        K\@. Werner\inst{1}
        \and
        P\@. H\@. Hauschildt\inst{2}
        \and
        J\@. W\@. Kruk\inst{3}
        }

\institute{Institute for Astronomy and Astrophysics,
           Kepler Center for Astro and Particle Physics,
           Eberhard Karls University,
           Sand 1,
           72076 T\"ubingen,
           Germany \\
           \email{rauch@astro.uni-tuebingen.de}
           \and
           Hamburger Sternwarte,
           Gojenbergsweg 112, 
           21029 Hamburg,
           Germany
           \and
           NASA Goddard Space Flight Center, 
           Greenbelt, 
           MD\,20771, 
           USA}

\date{Received 31 March 2015; accepted 25 April 2015}

\abstract {\aador is a close, totally eclipsing, post common-envelope binary with an sdOB-type primary star and an
           extremely low-mass secondary star,
           located close to the mass limit of stable central hydrogen burning. Within error limits, 
           it may  either be a brown dwarf or a late M-type dwarf.
          }
          {We aim to extract the secondary's contribution to the phase-dependent composite spectra.
           The spectrum and identified lines of the secondary decide on its nature.
          }
          {In January 2014, we measured the phase-dependent spectrum of \aador with XSHOOTER over one complete orbital period.
           Since the secondary's rotation is presumable synchronized with the orbital period, its surface  strictly divides  
           into a day and night side. Therefore, we may obtain the spectrum of its cool side during its transit
           and of its hot, irradiated side close to its occultation. We developed the Virtual Observatory (VO)
           tool TLISA to search for weak lines of a faint companion in a binary system. We successfully applied
           it to the observations of \aador.
          }
          {We identified 53 spectral lines of the secondary in the ultraviolet-blue, visual, and near-infrared XSHOOTER spectra
           that are strongest close to its occultation. We identified 57 (20 additional) lines in available UVES
           (Ultraviolet and Visual Echelle Spectrograph) spectra from 2001.
           The lines are mostly from \ion{C}{ii - iii} and \ion{O}{ii}, typical for a low-mass star that is 
           irradiated and heated by the primary.
           We verified the orbital period of $P = 22597.033201\pm 0.00007\,\mathrm{s}$ and 
           determined the orbital velocity
           $K_\mathrm{sec} = 232.9^{+16.6}_{\,\,-6.5}\,\mathrm{km/s}$
           of the secondary.
           The mass of the secondary is 
           $M_\mathrm{sec} = 0.081^{+0.018}_{-0.010}\,M_\odot$ 
           and, hence, it is not possible 
           to reliably determine a brown dwarf or an M-type dwarf nature.
          }
          {Although we identified many emission lines of the secondary's irradiated surface, 
           the resolution and signal-to-noise ratio
           of our UVES and XSHOOTER spectra are not good enough to extract 
           a good spectrum of the secondary's nonirradiated hemisphere.
          }

\maketitle

\section{Introduction}
\label{sect:intro}

\aador (\cp, \hd, \lb) was discovered to be a close, short-period, totally eclipsing binary \citep{kilkennyetal1978}
with an inclination of $i=89\fdg 21\pm 0\fdg 30$ \citep{hilditchetal2003}.
It is a post common-envelope system \citep{schreibergaensicke2003}
with a sdOB-type primary star and a low-mass companion \citep{rauch2004}.
Its orbital period $P = 0.261\,539\,7363 (4)$\,d (constant at a level of about $10^{-14}$\,d per orbit)
was determined by \citet{kilkenny2011} with high accuracy from the light curve, using eclipses from 1977 to 2010.

\aador belongs to the class of \hw-type variables.
These are eclipsing binaries consisting of a B-type subdwarf star and a late M-type star.
They experienced a common-envelope phase when the primary star evolved through a red-giant phase.
Their orbital periods are generally short \citep[$P \approx 0.1$\,d,][]{heber2009}.
They are of great importance because eclipsing binaries allow precise mass determinations.
Presently, we know more than 15 of these systems \citep{ritterkolb2003, almeidaetal2012, barlowetal2013,
schaffenrothetal2013, schaffenrothetal2014, schaffenrothetal2015}. 
Yet another sdOB-type member was discovered \citep[\object{NSVS\,14256825},][]{almeidaetal2012}.
\aador is of special importance because it is one of the hottest members with one of the longest periods.
It is one of the brightest ($m_\mathrm{V} = 11.138$)
and, hence, best-studied \hw-type binaries.

\aador is a precataclysmic variable \citep{ritter1986}
and, thus, stringent constraints for the mass of the low-mass secondary,
just at the limit of the hydrogen-burning mass, and its angular
momentum are a prerequisite for reliable common-envelope modeling,
especially the ejection mechanism \citep{liviosoker1984}
and predictions of the evolution of \aador.
Therefore, it is a key object for understanding common-envelope evolution.

Several spectral analyses by means of nonlocal thermodynamic equilibrium (NLTE) model atmospheres
were performed, and the primary star of \aador (\lb) turned out to be one of the hottest sdOB stars 
\citep[\Teffw{42000},][]{rauch2000,fleigetal2008}.
These analyses showed a discrepancy in the surface gravity 
that was derived by radial-velocity and light-curve analyses \citep{hilditchetal2003}, 
$\log (g\,/\,\mathrm{cm/s^2}) = 5.30 \pm 0.1$ and
$\log g = 5.53 \pm 0.03$, respectively.

Since \citet{vuckovicetal2008}  for the first time identified spectral lines of the secondary in spectra obtained with the
Ultraviolet and Visual Echelle Spectrograph 
(UVES\footnote{\url{http://www.eso.org/sci/facilities/paranal/instruments/uves.html}}) at the
European Southern Observatory (ESO)
and determined a lower limit ($K_\mathrm{sec} \ge 230 \pm 10\,$km/s) of its 
radial velocity amplitude, both components' masses are known
($M_\mathrm{pri} = 0.45$\,\Msol, $M_\mathrm{sec} = 0.076$\,\Msol),
however, with rather large error bars. 
\citet{muelleretal2010} considered that, because of the refection effect,
this measured $K_\mathrm{sec}$ comes from the outer edge
of the stellar disk. They applied the stellar radius from \citet[$R_\mathrm{sec} = 0.135\,R_\odot$]{rauch2000}
to shift the determined emission radius to the stellar center and used $K_\mathrm{sec} = 240\pm 20\,$km/s
to calculate the stellar masses.
With $K_\mathrm{pri} = 40.15\pm 0.11\,$km/s, they derived
$M_\mathrm{pri} = 0.51^{+0.125}_{-0.108}$\,\Msol and
$M_\mathrm{sec} = 0.085^{+0.031}_{-0.023}$\,\Msol.

\begin{figure}
   \resizebox{\hsize}{!}{\includegraphics{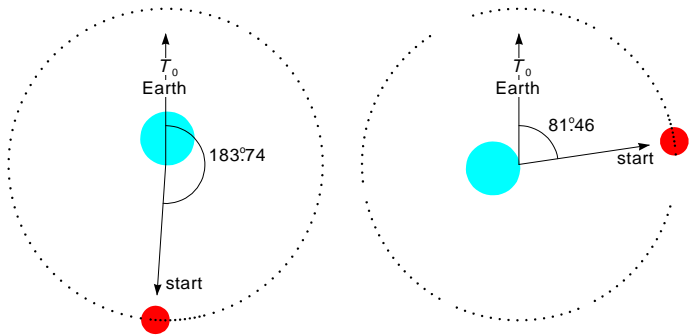}}
    \caption{Orbital location of the primary (light blue in the online version) and secondary (red) at the 
             UVES (left) and XSHOOTER (right) observation times. The start times  of the two observation
             campaigns are indicated  (HJD 2451917.5191833 and
             2456665.5858551, respectively). 
             The stars move counterclockwise.
             Dots mark the locations of the secondary at the start times of the individual observations. 
            }
   \label{fig:obstimes}
\end{figure}

We  recently performed an NLTE spectral analysis of the primary \citep{klepprauch2011},
where we used a lower rotational velocity of $v_\mathrm{rot} = 30\,$km/s 
\citep{muelleretal2010,geieretal2010}
than before
\citep[35\,km/s,][]{fleigetal2008}.
With improved model atmospheres and improved Stark line-broadening
tables for H\,{\sc i} lines \citep{tremblaybergeron2009},
we determined
$T_\mathrm{eff}\hspace{-0.5mm}=\hspace{-0.5mm}42000\pm 1000\,\mathrm{K}$ and  
\mbox{$\log g\hspace{-0.5mm} =\hspace{-0.5mm} 5.46\pm 0.05$}.
Thus, the gravity problem is solved.
From our mass determination of $M_\mathrm{pri} = 0.4714\pm 0.0050$\,\Msol,
in comparison with evolutionary tracks of post-extended horizontal branch (post-EHB) stars
\citep{dormanetal1993},
we calculated ($M_\mathrm{pri}K_\mathrm{pri} = M_\mathrm{sec}K_\mathrm{sec}$) 
the secondary's mass to be $M_\mathrm{sec} = 0.0725 - 0.0863$\,\Msol.
Since the hydrogen-burning mass limit is about $0.075$\,\Msol 
\citep{chabrierbaraffe1997,charbrieretal2000},
the secondary may be either a brown dwarf or a late M-type dwarf.

To make progress in our understanding of common-envelope evolution in general
and to further constrain the secondary's nature, 
we measure the secondary's contribution to the
composite spectrum of \aador, especially in the infrared and follow 
its spectral evolution during one complete orbital period.
We therefore performed phase-dependent spectroscopy with 
XSHOOTER\footnote{\url{http://www.eso.org/sci/facilities/paranal/instruments/xshooter.html}.} \citep{vernetetal2011}
since this contribution is of the order of a few percent
\citep{fleigetal2008} and, thus, a very high
signal-to-noise ratio (S/N) of the observed spectra is required for its detection.

In Sect.\,\ref{sect:observation}, we briefly describe our new XSHOOTER observations and
the UVES data used by \citet{rauchwerner2003}. 
We determine the orbital period in Sect.\,\ref{sect:period}.
We then revisit the sdOB primary (Sect.\,\ref{sect:primary})
and demonstrate how the recently determined surface gravity of \citet{klepprauch2011} impacts the
element abundance determination of \citet{fleigetal2008}. 
In Sect.\,\ref{sect:secondary}, we turn to the
secondary star of \aador and introduce the newly developed, registered Virtual Observatory (VO) tool
T\"ubingen Line Identification and Spectrum Analyzer (TLISA),
which is part of the T\"ubingen German Astrophysical Virtual Observatory
(GAVO\footnote{\url{http://www.g-vo.org}})  
project. Its application to the XSHOOTER and UVES spectra of \aador is described.
In Sect.\,\ref{sect:masses}, we present our results for the orbit dimensions of \aador, 
as well as the masses and radii of its components. 
We summarize and conclude in Sect.\,\ref{sect:results}.

\section{Observations}
\label{sect:observation}

We took 105 UVES spectra (individual exposure times 180\,s, $3743\,\mathrm{\AA} - 4986\,\mathrm{\AA}$), 
which cover a complete orbital period ($P = 0.26\,\mathrm{d}$), 
on Jan 8, 2001 (program 066.D-1800) at the 
Very Large Telescope / Unit Telescope 2 (VLT/UT2, Kueyen).
The spectra were subject to the standard reduction provided by ESO.
With a slit width of $1\farcs0$ projected at the sky,
a resolving power $R \approx 48000$ was achieved. The S/N of the single spectra 
is about $20\,-\,30$. See \citet{rauchwerner2003} for details.

We obtained 300 XSHOOTER spectra (in its three arms,
UVB\footnote{ultraviolet-blue, $3000\,\mathrm{\AA} - 5595\,\mathrm{\AA}$, $R = 4350$}, 
VIS\footnote{visual, $5595\,\mathrm{\AA} - 10240\,\mathrm{\AA}$, $R = 7450$}, and 
NIR\footnote{near infrared, $10240\,\mathrm{\AA} - 24800\,\mathrm{\AA}$, $R = 5300$},
each with 100 exposures of 
160\,s,
170\,s, and
34\,s, respectively)
 exactly 13 years later on Jan 8, 2014 (program 092.C-0692) at the 
VLT/UT3 (Melipal). The observations of \aador
started at 02:03 UT (air mass 1.4) and ended at 08:31 UT
(air mass 2.3), just before morning twilight. 
The seeing\footnote{\url{http://www.ls.eso.org/lasilla/dimm/Archive}} 
was around 1\arcsec\ for much of the night but deteriorated greatly during the last two observation
blocks (OBs), reaching almost 3\arcsec\ for the final OB (Roger Wesson, priv\@. comm.). 
We reduced this data with ESO's Reflex Environment\footnote{\url{https://www.eso.org/sci/software/reflex}}
(version 2.6, downloaded Feb 13, 2015, XSHOOTER pipeline identifier 2.5.2). The S/N of the spectra is about $30\,-\,50$.

Figure\,\ref{fig:obstimes} displays the binary phases at the start times of the individual UVES and XSHOOTER observations.
Figures\,\ref{fig:uves} -- \ref{fig:nir} show each  individual spectrum
of the UVES and XSHOOTER observation campaigns. The XSHOOTER spectra are compared with our best model for the primary
(see Sect.\,\ref{sect:primary}). The synthetic spectrum is normalized to match the
2MASS $\mathrm{K_s} = 12.046$ brightness. A reddening with $\ebv = 0.08$ is applied following the law of 
\citet[][with $R_V = 3.1$]{fitzpatrick1999}. The synthetic spectrum
is convolved with a Gaussian to match the XSHOOTER resolution.

\onlfig{
\begin{landscape}
\begin{figure*}
   \includegraphics[trim=0 200 0 0,width=0.9\textwidth,angle=270]{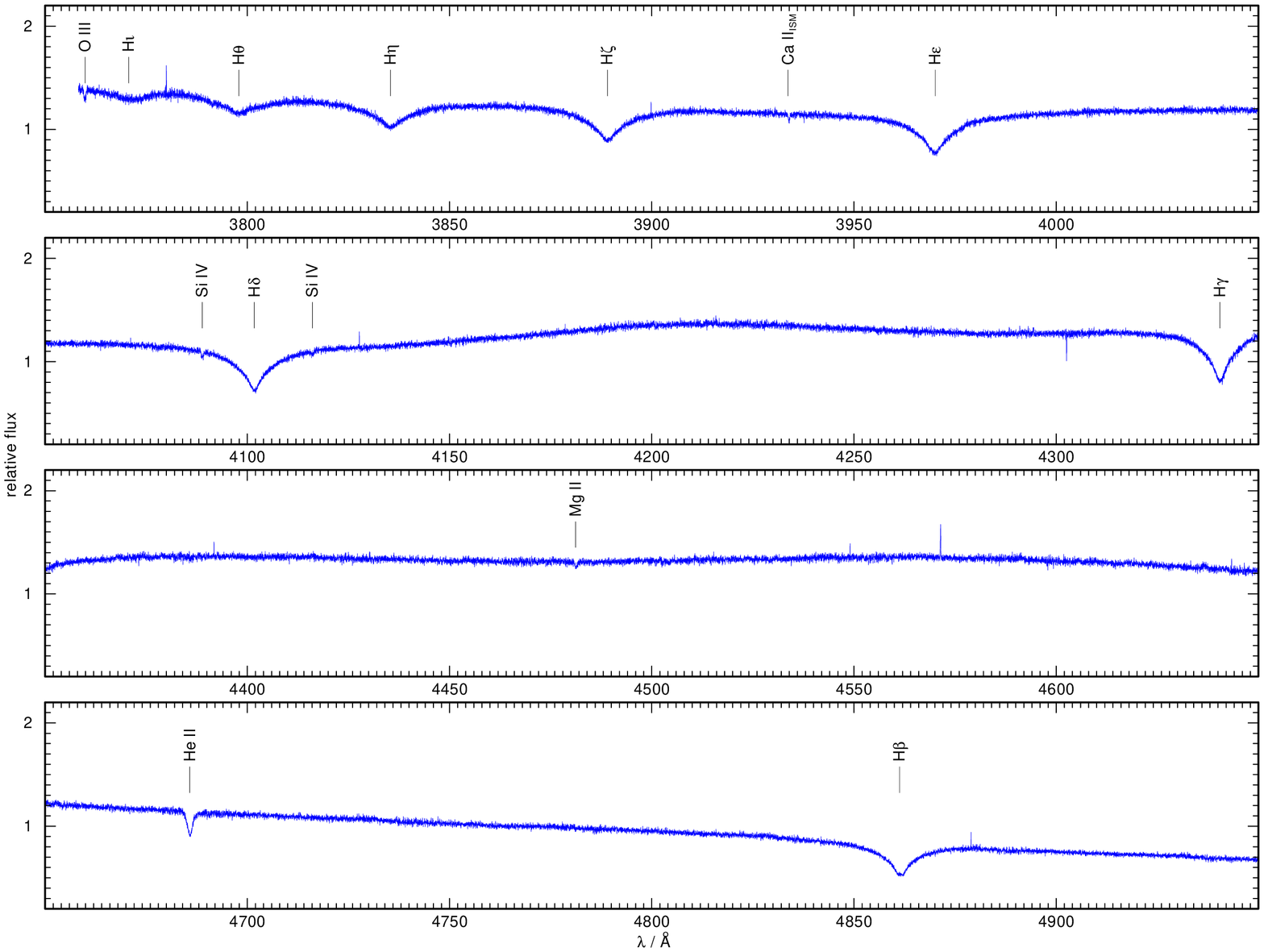}
    \caption{UVES spectrum (blue, start time 2001-01-08T00-27-37.434 UT). \newline
             Identified lines of the primary are marked. ISM denotes interstellar lines.
            }
   \label{fig:uves}
\end{figure*}
\end{landscape}
}

\onlfig{
\begin{landscape}
\begin{figure*}
   \includegraphics[trim=0 200 0 0,width=0.9\textwidth,angle=270]{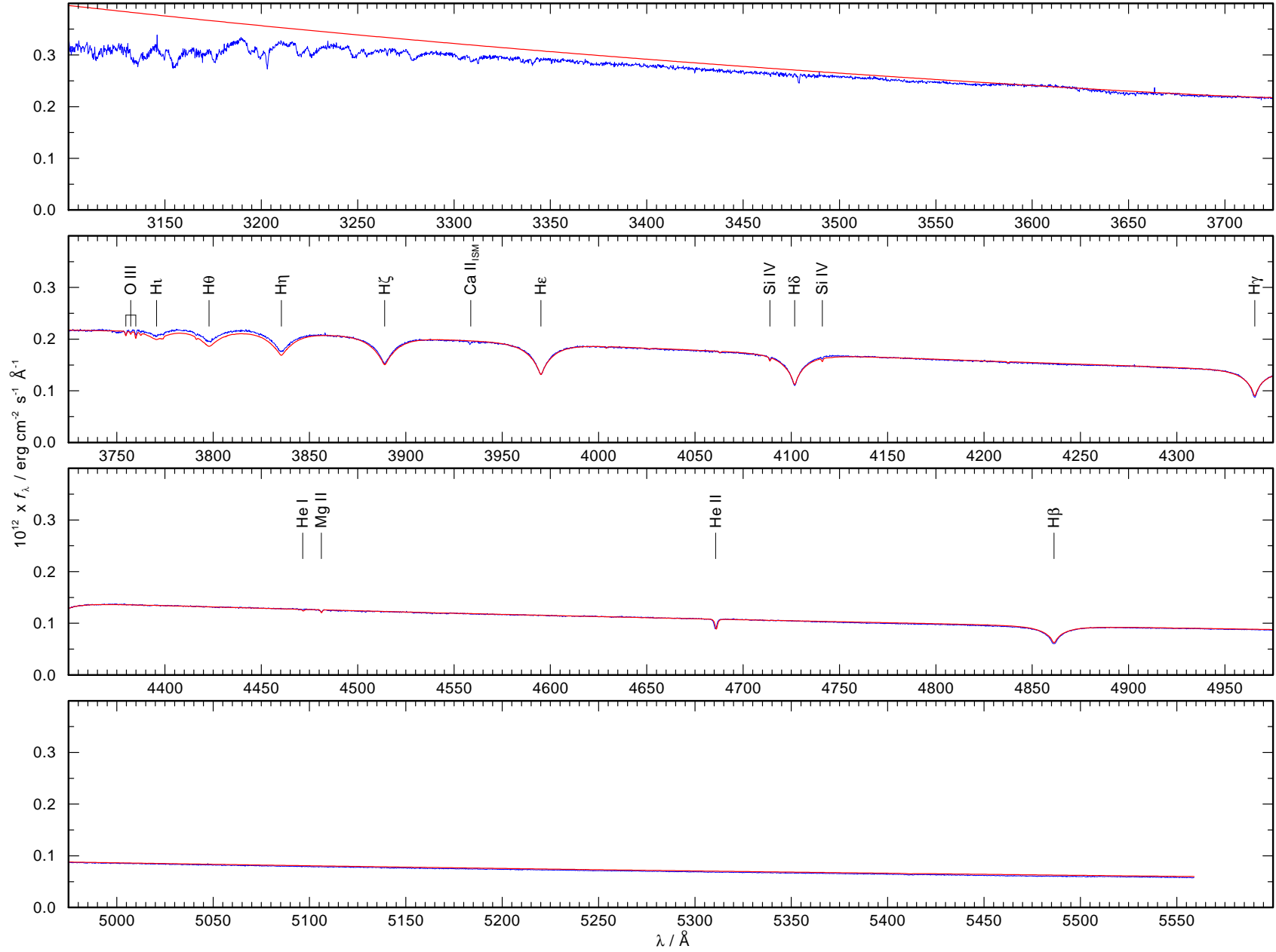}
    \caption{XSHOOTER spectrum (blue, start time 2014-01-08T02-03-37.877 UT) taken in the UVB arm compared \newline
             with our final model of the primary (red).
             Identified lines of the primary are marked.  ISM denotes interstellar lines.
            }
   \label{fig:uvb}
\end{figure*}
\end{landscape}
}

\onlfig{
\begin{landscape}
\begin{figure*}
   \includegraphics[trim=0 200 0 0,width=0.9\textwidth,angle=270]{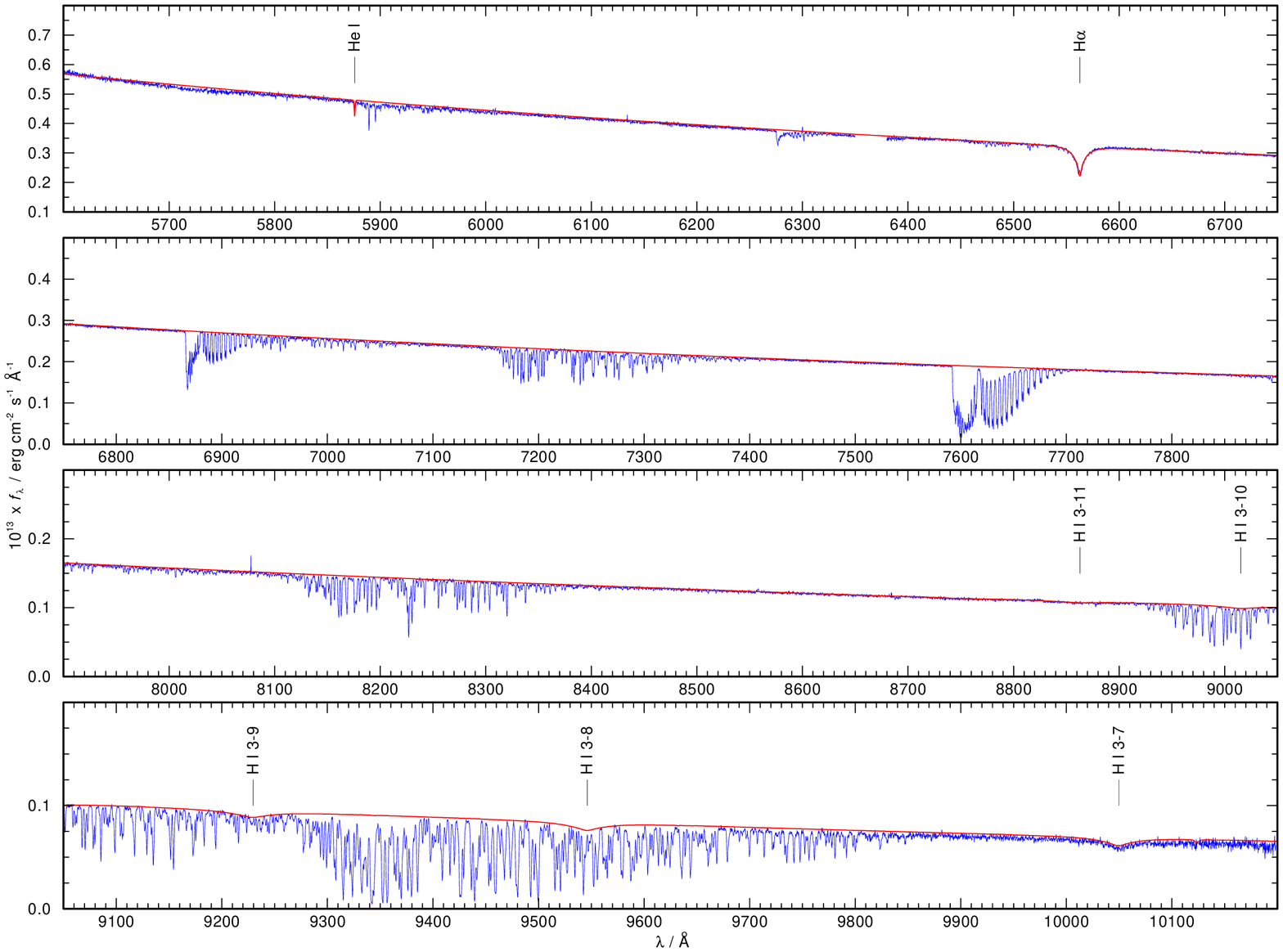}
    \caption{XSHOOTER spectrum (blue, start time 2014-01-08T02-03-43.068 UT) taken in the VIS arm compared \newline
             with our final model of the primary (red).
             Identified lines of the primary are marked.
            }
   \label{fig:vis}
\end{figure*}
\end{landscape}
}

\onlfig{
\begin{landscape}
\begin{figure*}
   \includegraphics[trim=0 200 0 0,width=0.9\textwidth,angle=270]{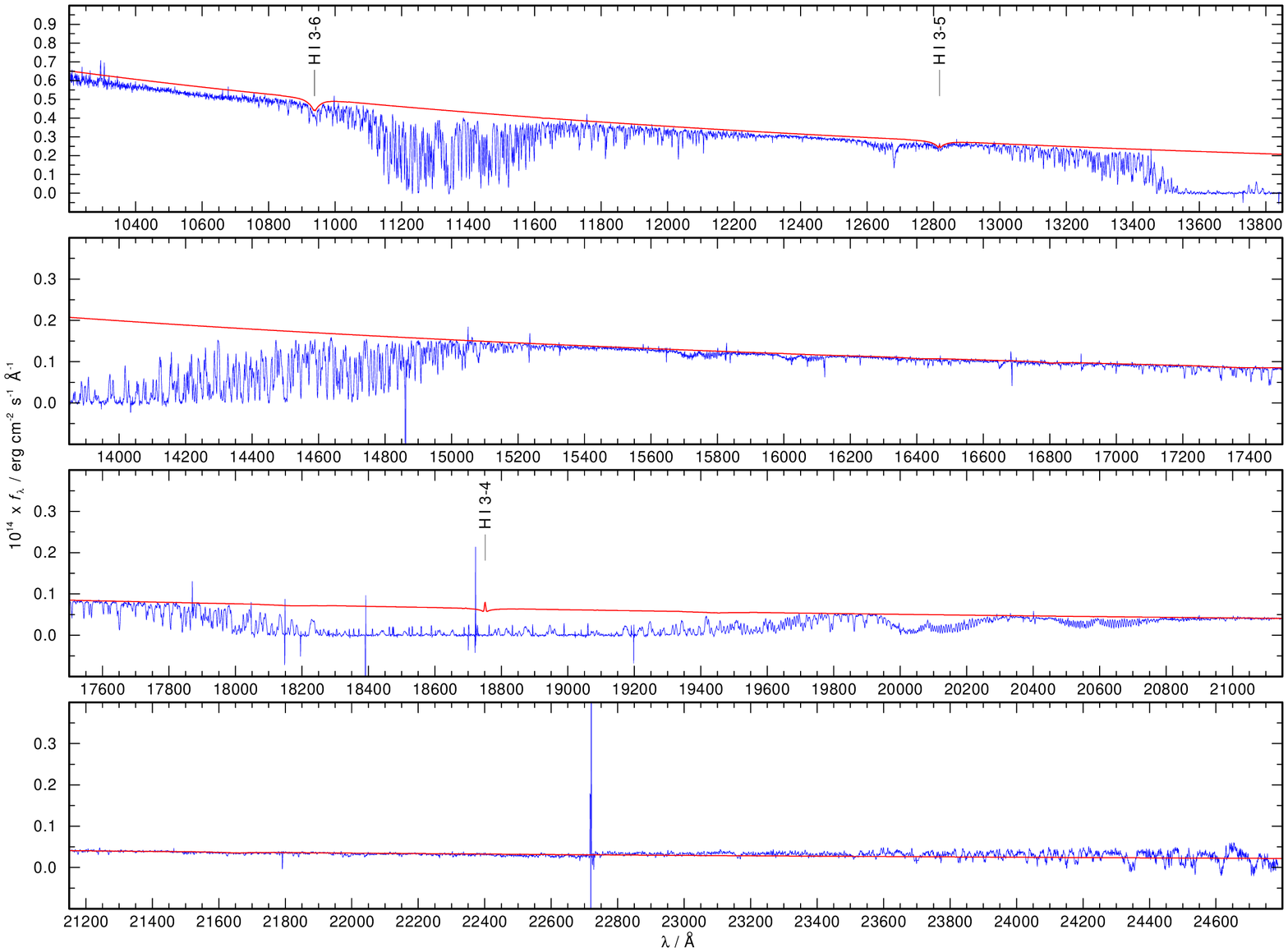}
    \caption{XSHOOTER spectrum (blue, start time 2014-01-08T02-03-45.7348) UT) taken in the NIR arm compared \newline
             with our final model of the primary (red).
             Identified lines of the primary are marked.
            }
   \label{fig:nir}
\end{figure*}
\end{landscape}
}

\section{Orbital period}
\label{sect:period}

\citet{rauchwerner2003} determined an orbital period of $P = 0.261\,582\,199 (58)\,\mathrm{d}$
from the \ionw{He}{ii}{4686} radial-velocity curve.
This deviates ($\approx +0.016$\,\%) from that of \citet[$P = 0.261\,539\,7363 (4)\,\mathrm{d}$]{kilkenny2011}
measured from the light curve. Our XSHOOTER data extended the timebase of the 
radial-velocity curve from 2535 days to 7283 days 
\citep[including the measurements of][and of FUSE observations in 2003 and 2004]{hilditchetal1996,rauchwerner2003}.
We employed the Interactive Data Language
(IDL\footnote{\url{http://www.exelisvis.com/ProductsServices/IDL}}) to measure the Doppler shifts
of \ionw{He}{ii}{4686} in the optical
and determined those of 
\ion{C}{iii}, \ion{O}{vi}, \ion{Si}{iv}, and \ion{P}{v} lines in the far-ultraviolet (FUSE).

\begin{figure}
   \resizebox{\hsize}{!}{\includegraphics{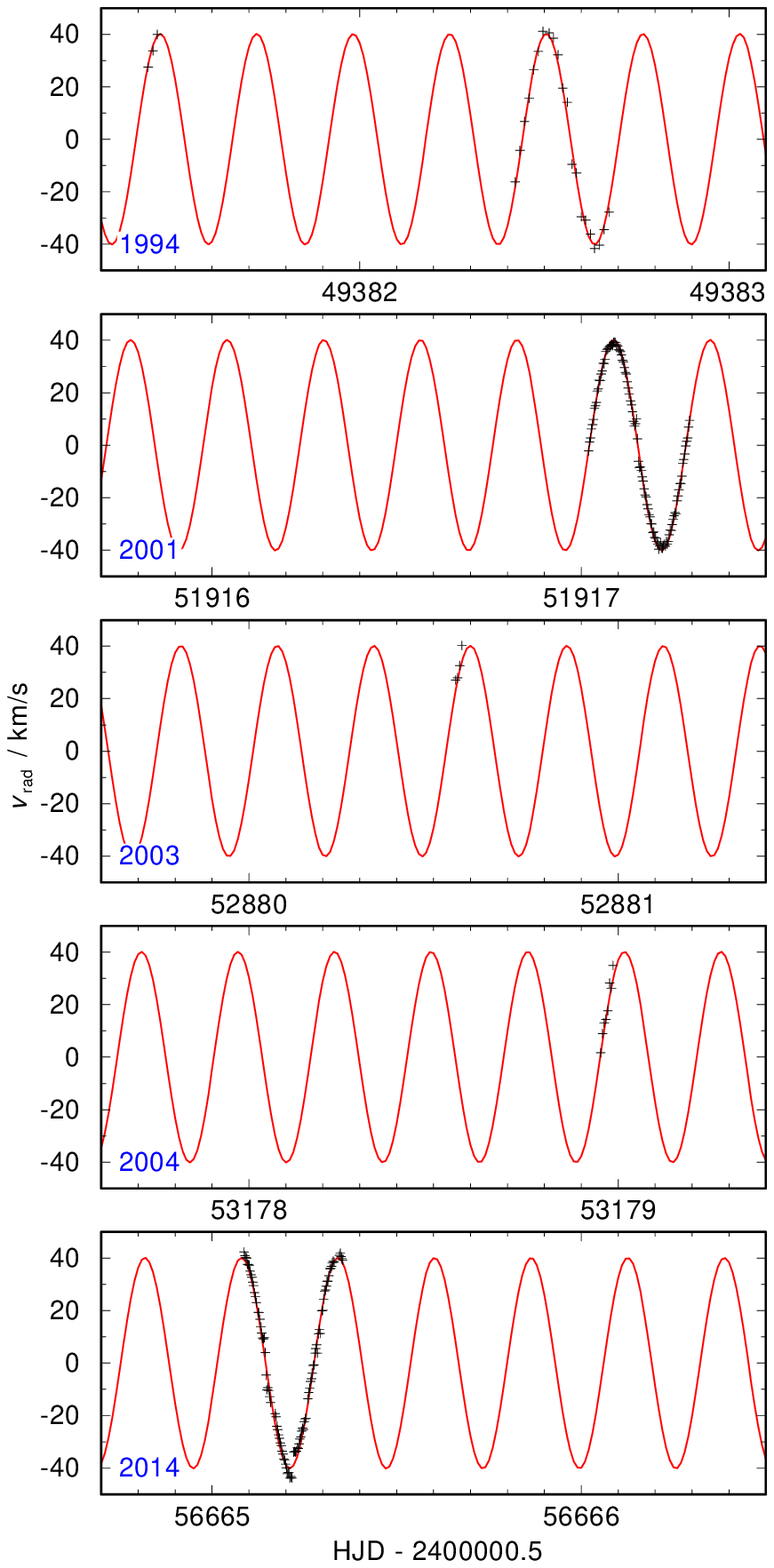}}
    \caption{Comparison of radial-velocity measurements 
             of \aador with a sine curve
             calculated with the period of 
             $P = 0.261\,539\,7363$\,d 
             from the light-curve solution of
             \citet{kilkenny2011}.
             $K_\mathrm{pri}=40.15\,\mathrm{km/s}$ 
             and
             $T_\mathrm{0} = 51917.15269$ are from 
             \citet{muelleretal2010} and \citet{rauchwerner2003}, respectively.
             From top to bottom, the panels show data from
             \citet{hilditchetal1996},
             \citet[UVES]{rauchwerner2003},
             FUSE 2003,
             FUSE 2004, 
             and our XSHOOTER spectra.
             The measurements cover the time from Jan 29, 1994, to Jan 8, 2014.
            }
   \label{fig:period}
\end{figure}

We used the TRIPP software package \citep{schuhetal2003,geckeleretal2014} to determine the
period \citep[with $T_\mathrm{0} = 2451917.15269$, the time of mid-transit,][]{rauchwerner2003}.
We confirmed Kilkenny's light-curve solution 
(Fig.\,\ref{fig:period}, $P = 22597.03320768 \pm 0.0000691\,\mathrm{s}$, Sect.\,\ref{sect:intro}).

We note that Fig.\,4 of \citet{rauchwerner2003} is incorrect. 
In its top panel,
the data of \citet{hilditchetal1996} is erroneously shown with
an offset of +0.5\,d, and the $v_\mathrm{rad}$ values in the bottom panel have
an incorrect sign.
This explains the period deviation that was previously found (Sect.\,\ref{sect:intro}).

\section{The primary}
\label{sect:primary}

To disentangle the composite \aador spectra and to extract the secondary's spectrum,
we must subtract the primary's contribution, considering its orbital motion, from all
spectra. Therefore, we shifted all spectra to
rest wavelengths, i.e., we applied a phase-dependent radial-velocity correction for the primary.
Then, we coadded the spectra with the best S/N ratio to achieve a master spectrum.
The expected weak lines of the secondary smear out in this procedure and leave no detectable marks.

Since the primary's master spectra (separately for UVES and XSHOOTER) allowed us to verify the result of 
\cite{klepprauch2011} for the surface gravity (\loggw{5.46\pm 0.05}), we decided to calculate a new grid of NLTE
atmosphere models with our T\"ubingen NLTE Model-Atmosphere Package
\citep[TMAP\footnote{\url{http://astro.uni-tuebingen.de/~TMAP}};][]{werneretal2003,rauchdeetjen2003,tmap2012}
to investigate  the impact of the higher \logg on the photospheric abundances
that were determined by \citet{fleigetal2008}, who used \loggw{5.3}.
An indication that these abundances may be slightly different is the prominent
Balmer-line problem \citep{napiwotzkirauch1994,rauch2000,rauch2012a} due to the underestimation of metal opacities 
\citep{bergeronetal1993,werner1996}. The effect is apparently stronger at the
higher \logg \citep[][their Fig.\,5]{klepprauch2011}. Therefore the abundance
determination was repeated. 
The H, He, C, N, O, Mg, Si, P, S, Fe, and Ni abundances
were determined by spectral fits to ultraviolet observations
obtained with the International Ultraviolet Explorer \citep[IUE, cf\@.][]{rauch2000} and
the Far Ultraviolet Spectroscopic Explorer \citep[FUSE, cf\@.][]{fleigetal2008}
and to our UVES and XSHOOTER spectra. Table\,\ref{tab:abund} shows the adjusted abundance values.
The errors are typically $\pm 0.2$\,dex.
Significant deviations were found for Mg, Si, P, and Ni. They are not due to the higher \logg but
to an improved spectral fitting.

\begin{table}\centering
\caption{Photospheric abundances (mass fractions) of the primary of \aador, determined by 
\citet[][\Teffw{42000}, \loggw{5.30}, denoted F]{fleigetal2008} and 
in this work (H, \Teffw{42000}, \loggw{5.46}). Column 4 shows the deviations.}
\label{tab:abund}
\begin{tabular}{llr@{.}lr@{.}lrl}
\hline
\hline
\multicolumn{2}{c}{Element} & \multicolumn{2}{c}{F} & \multicolumn{2}{c}{H} & \multicolumn{2}{c}{(H/F)$-1$ / \%} \\
\hline
\noalign{\smallskip}
& H  & $9$&$9 \times 10^{-1}$ & $9$&$9 \times 10^{-1}$ & $   0$ & \\
& He & $3$&$2 \times 10^{-3}$ & $2$&$7 \times 10^{-3}$ & $ -15$ & \\
& C  & $1$&$8 \times 10^{-5}$ & $2$&$0 \times 10^{-5}$ & $  10$ & \\
& N  & $4$&$1 \times 10^{-5}$ & $4$&$6 \times 10^{-5}$ & $  10$ & \\
& O  & $1$&$0 \times 10^{-3}$ & $1$&$1 \times 10^{-3}$ & $  10$ & \\
& Mg & $3$&$6 \times 10^{-3}$ & $4$&$1 \times 10^{-4}$ & $ -90$ & \\
& Si & $9$&$7 \times 10^{-5}$ & $2$&$1 \times 10^{-4}$ & \hbox{}\hspace{5mm}$ 110$ & \\
& P  & $5$&$2 \times 10^{-6}$ & $2$&$6 \times 10^{-6}$ & $ -50$ & \\
& S  & $3$&$2 \times 10^{-6}$ & $3$&$2 \times 10^{-4}$ & $ 100$ & \\
& Fe & $1$&$2 \times 10^{-3}$ & $1$&$2 \times 10^{-3}$ & $   0$ & \\
& Ni & $3$&$5 \times 10^{-4}$ & $5$&$2 \times 10^{-4}$ & $  50$ & \\
\hline
\end{tabular}
\end{table}

\begin{figure}
   \resizebox{\hsize}{!}{\includegraphics{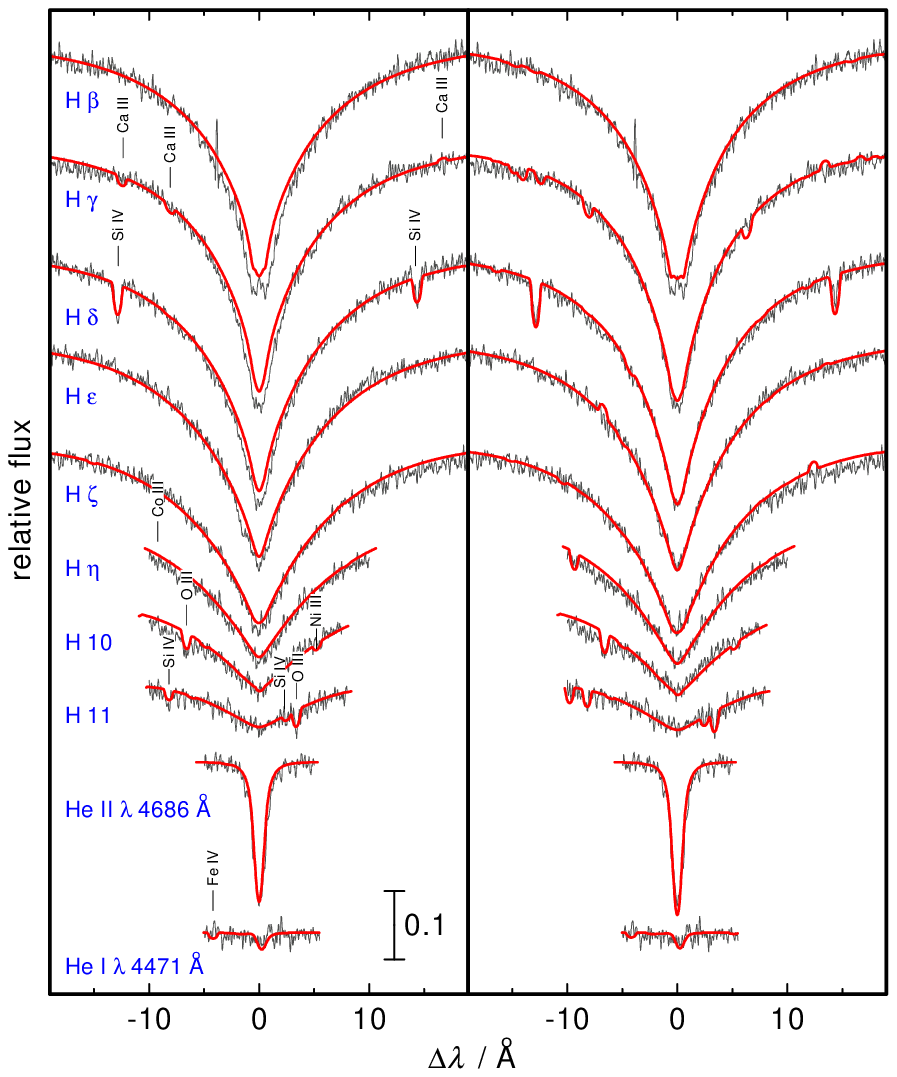}}
    \caption{Comparison of synthetic profiles of H and He lines calculated from our models (\Teffw{42000}, \loggw{5.46})
             with our UVES observation \citep[for clarity smoothed with a low-pass filter,][]{savitzkygolay1964}.
             Left panel: abundances from Table\,\ref{tab:abund}, and Sc, Ti, V, Cr, Mn, and Co with
             solar abundances \citep{scottetal2015b}. Right panel: similar model in the left panel, except the
             Sc, Ti, V, Cr, Mn, and Co abundances are artifically increased by a factor of 700.
             The vertical bar indicates 10\,\% of the continuum flux.}
   \label{fig:bergeron}
\end{figure}

The Balmer-line problem, however, is still present at the same strength with the new, fine-tuned abundances.
An additional model-atmosphere calculation with artificially increased 
\citep[to 700 times solar,][]{scottetal2015b,scottetal2015a} 
iron-group (here Ca -- Ni) abundances shows that the Balmer-line problem almost vanishes (Fig.\,\ref{fig:bergeron}). 
This corroborates the suggestion that missing metal opacities and, hence, cooling of the outer atmosphere are 
in general responsible  for the Balmer-line problem. The real opacity source in \aador, however, is not known yet.
On the one hand, the rotation of its primary \citep[$v_\mathrm{rot} = 30\,$km/s, cf\@.][]{klepprauch2011}
prohibits the detection of weak metal lines, and on the other hand, trans-iron-group elements
\citep[e.g.,][]{werneretal2012,rauchetal2012ge,rauchetal2014zn,rauchetal2014ba,rauchetal2015ga} have been totally neglected so far.

\section{The secondary}
\label{sect:secondary}

\begin{figure}
   \resizebox{\hsize}{!}{\includegraphics{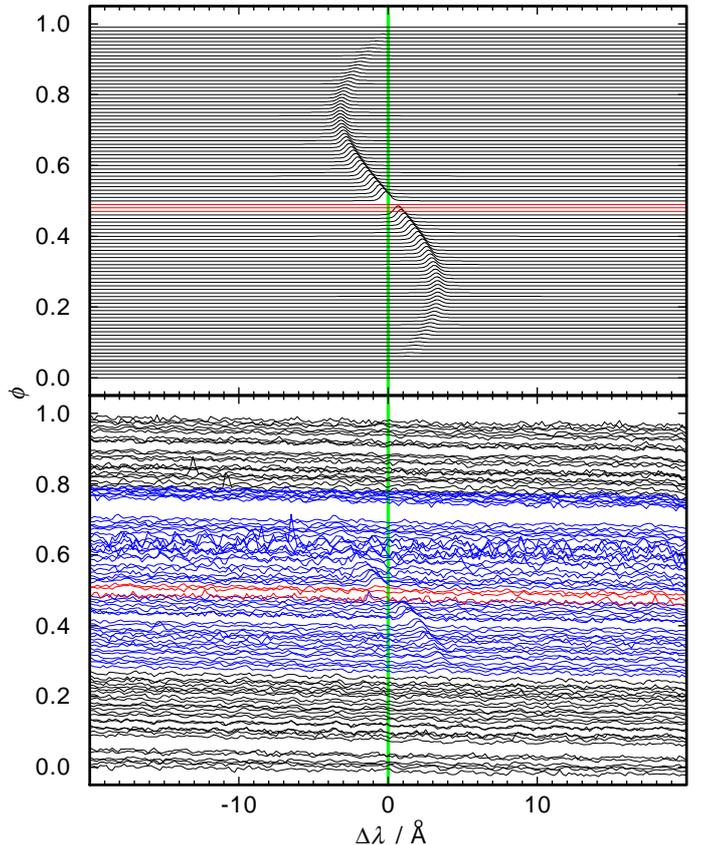}}
    \caption{Top: \Jonw{C}{ii}{4267.09} line in a simulation (with an assumed $K_\mathrm{sec} = 230$\,km/s).
             $\phi$ is the orbital phase. The transit ($\phi = 0.0$) and the occultation ($\phi = 0.5$)
             of the secondary are simulated.
             Bottom: the same line in our XSHOOTER observations (bottom). In the online version, the red spectra
             show the occultation of the secondary. In the blue spectra,
             \Jonw{C}{ii}{4267.09} is prominent. The vertical (green) bar indicates the
             rest wavelength of the line.
            }
   \label{fig:scurve}
\end{figure}

Binary systems often reveal themselves by the sinusoidal movement of their components' lines
in trailed spectra \citep[e.g.,][]{maxtedetal2000}. Figure\,\ref{fig:scurve} shows
a comparison of a synthetic and an observed ``s-curve''. 
To search for lines of the \aador secondary,
the primary's master spectra (Sect.\,\ref{sect:primary}) are subtracted from the 
UVES and XSHOOTER observations with a respective phase-dependent radial-velocity correction.

To make the application of the s-curve method easier, 
we have created a tool (Sect.\,\ref{sect:tlisa}) that even detects  weak s-curves and that measures the orbital period and 
velocity amplitude via this sensitive visual method.

\subsection{The new GAVO tool TLISA}
\label{sect:tlisa}

In the framework of the T\"ubingen GAVO project, we  developed the
T\"ubingen Line Identification and Spectrum Analyzer
(TLISA) tool that evaluates spectra and visualizes them. The first
module, SEarch for Faint Objects in Binary Systems (SEFOBS), allows us to
correct for orbital phase and velocity amplitude.
Its sensitivity, i.e., the detection limit for lines is demonstrated in Fig.\,\ref{fig:detlimit}.

\onlfig{
\begin{figure*}
   \includegraphics[trim=0 0 0 0,height=6.40CM,angle=0]{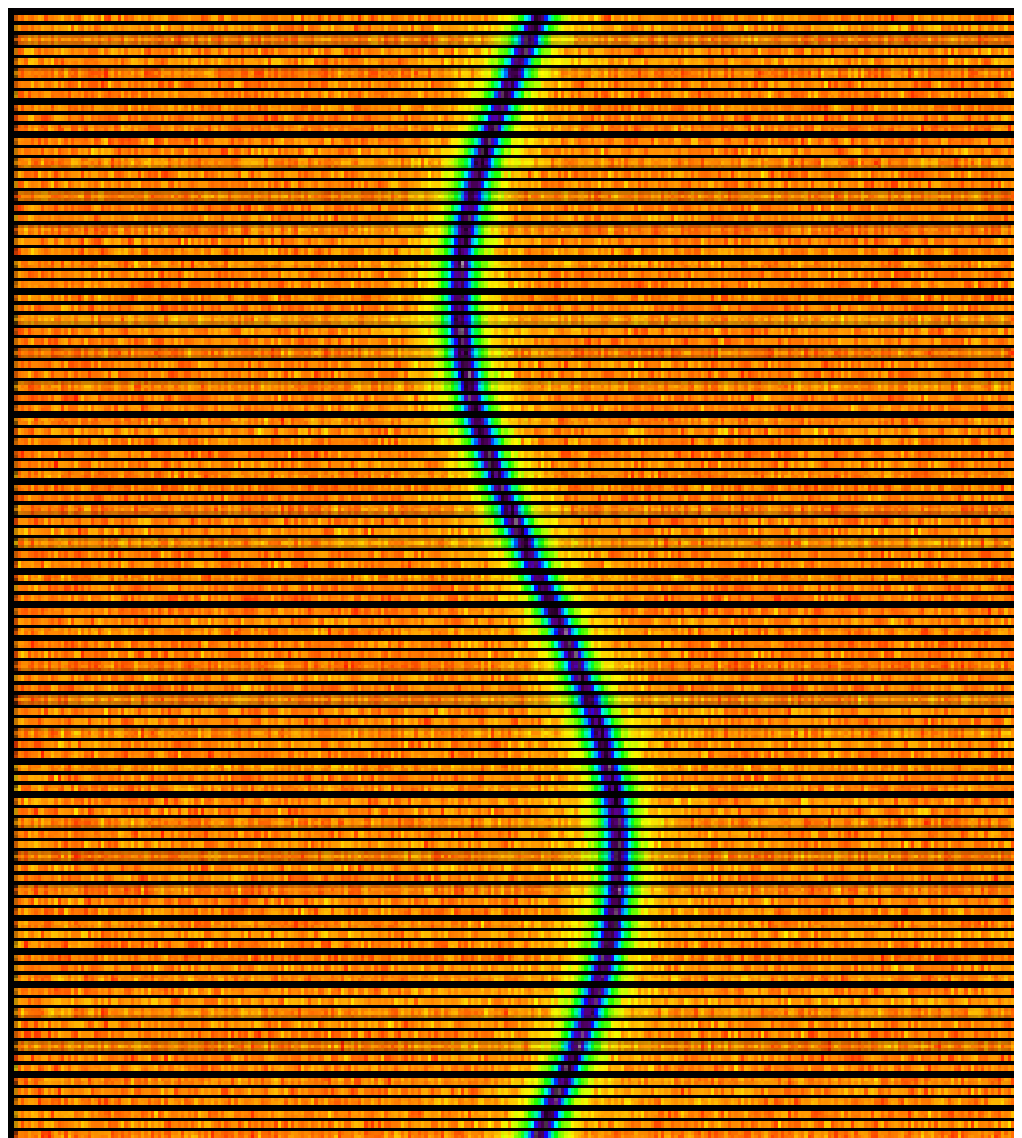}
   \includegraphics[trim=0 0 0 0,height=6.40CM,angle=0]{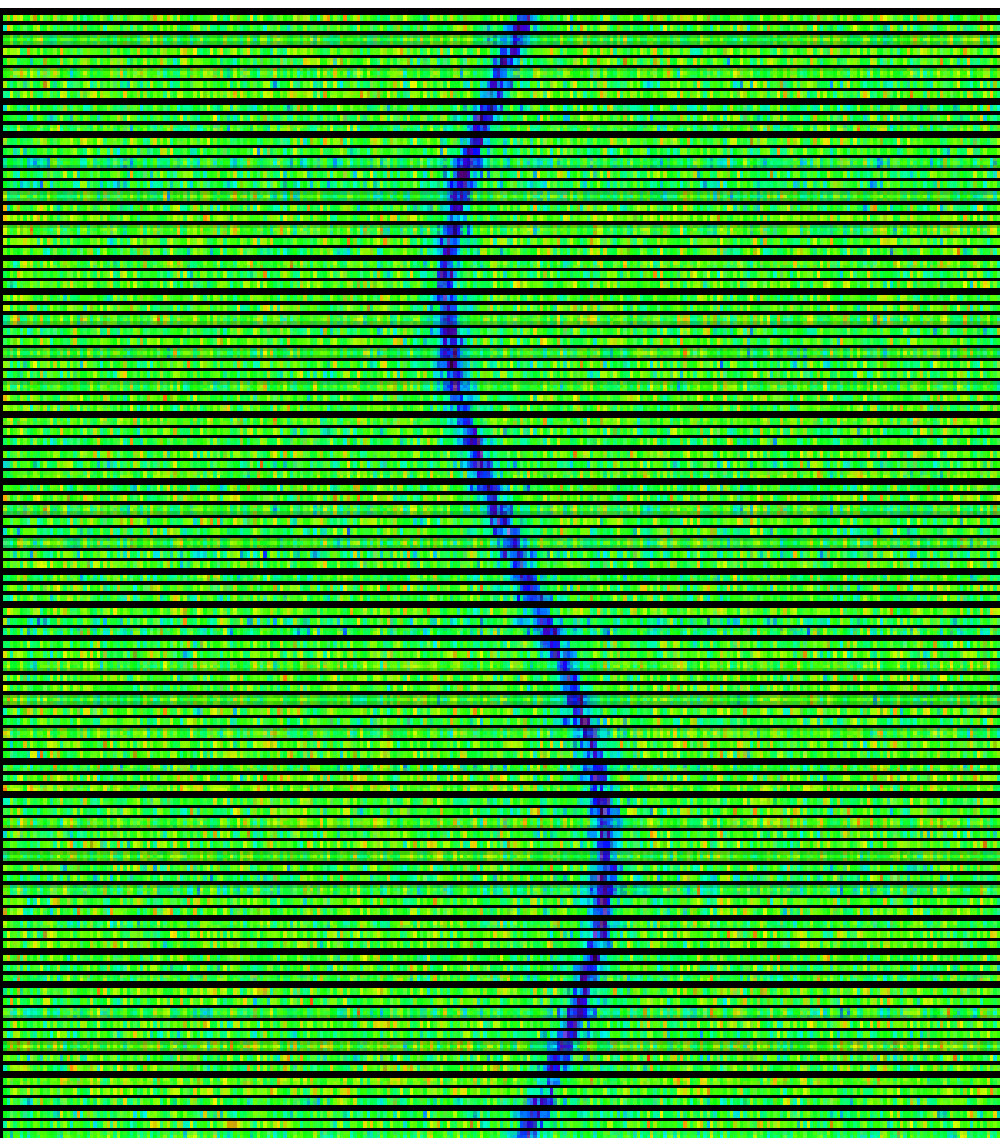}
   \includegraphics[trim=0 0 0 0,height=6.40CM,angle=0]{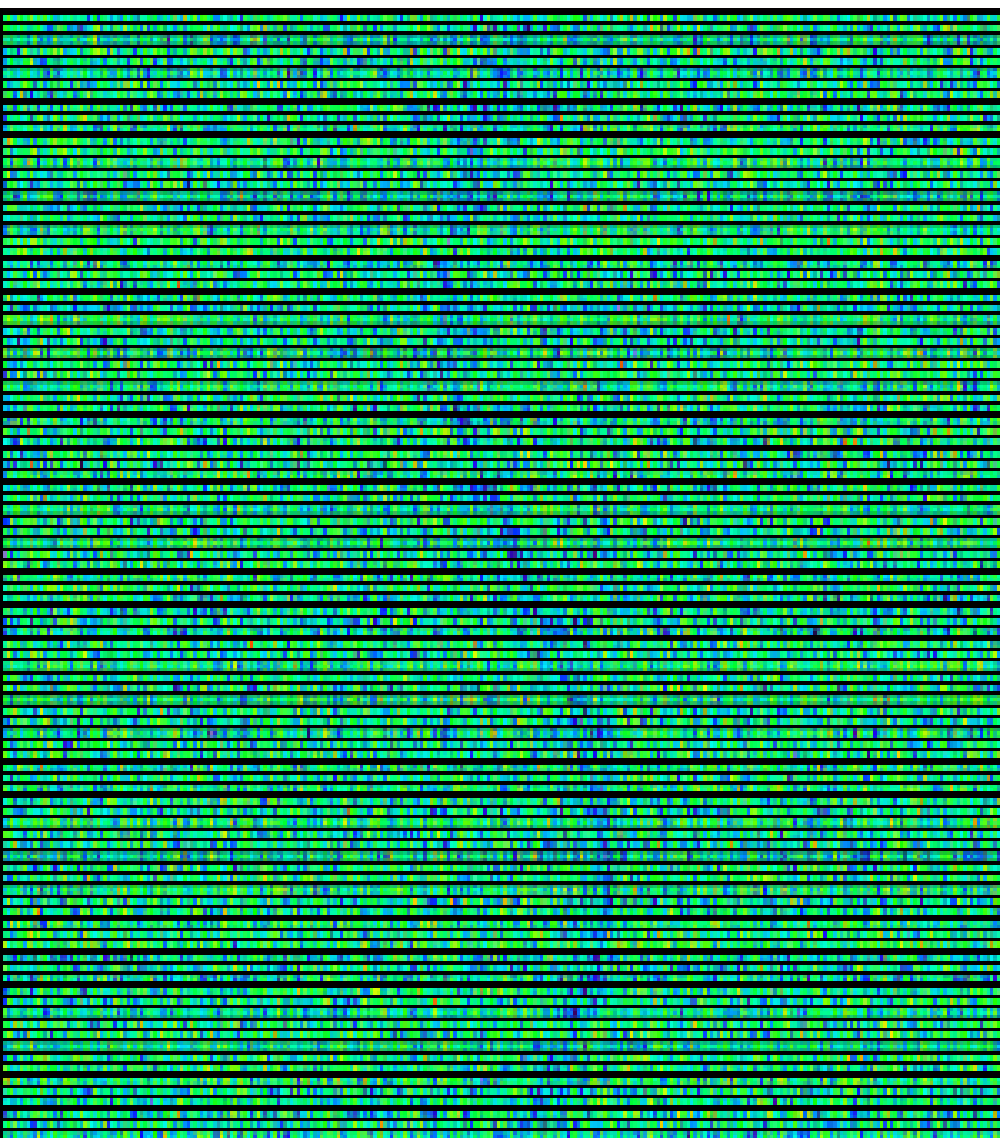}
   \includegraphics[trim=0 0 0 0,height=6.40CM,angle=0]{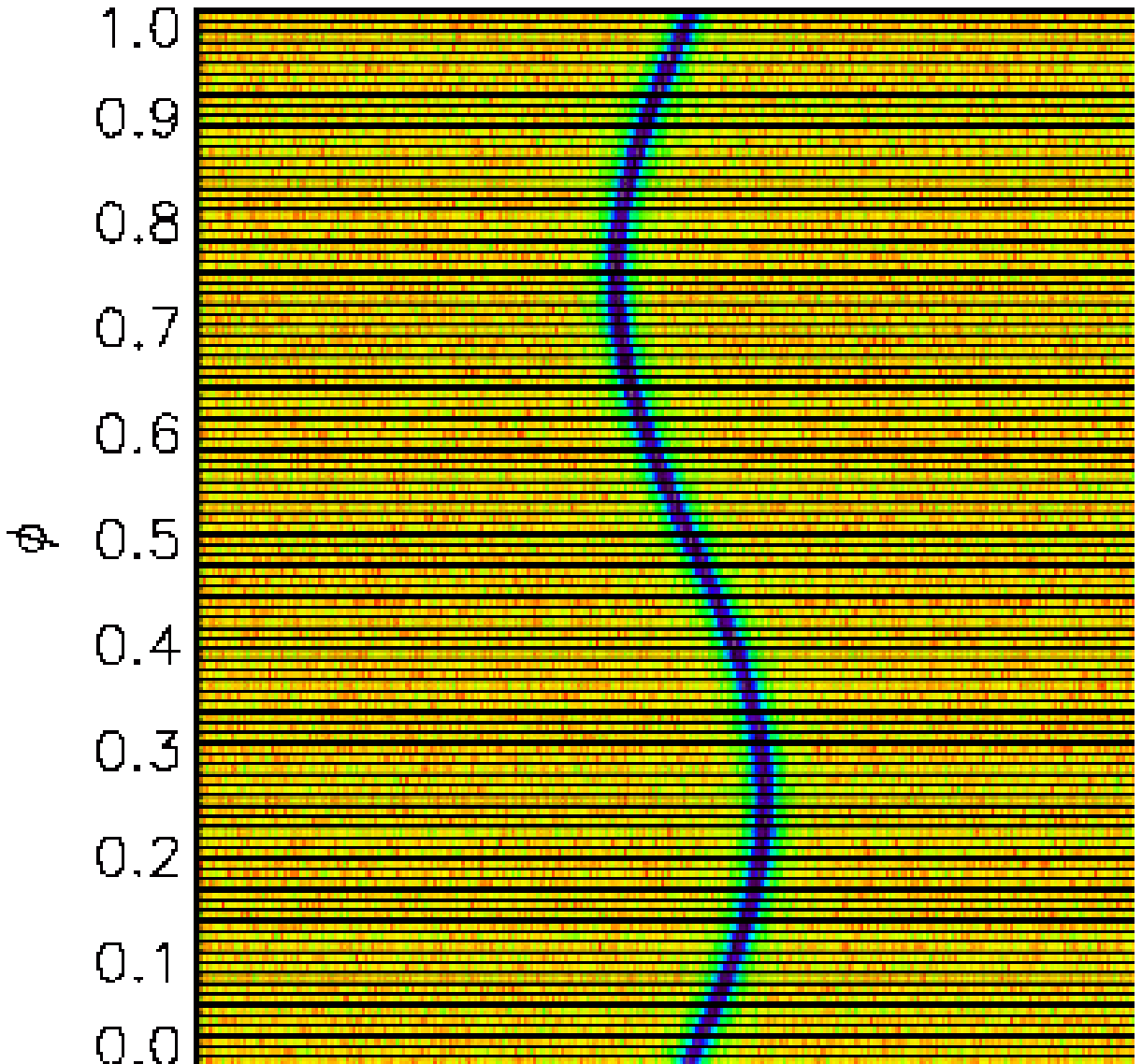}
   \includegraphics[trim=0 0 0 0,height=6.40CM,angle=0]{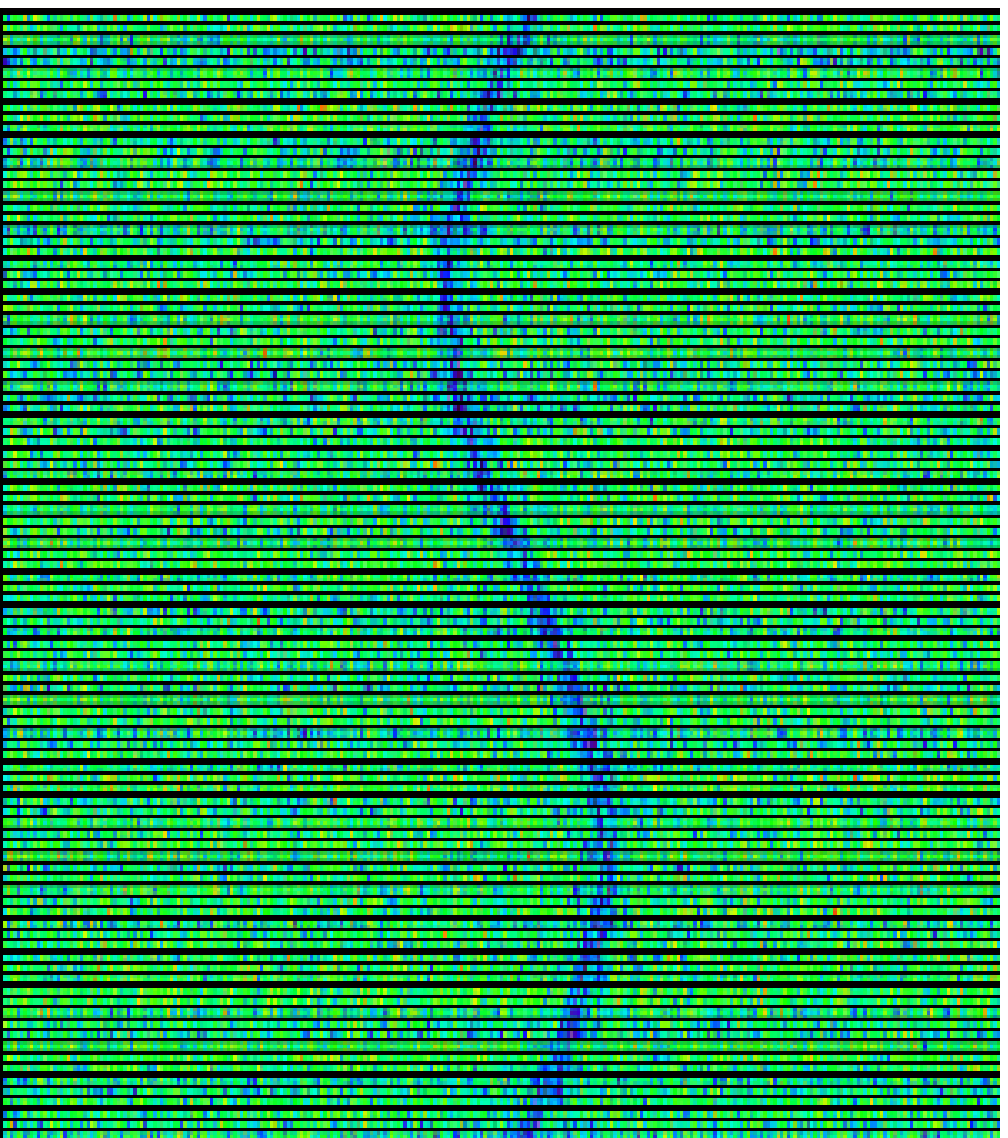}
   \includegraphics[trim=0 0 0 0,height=6.40CM,angle=0]{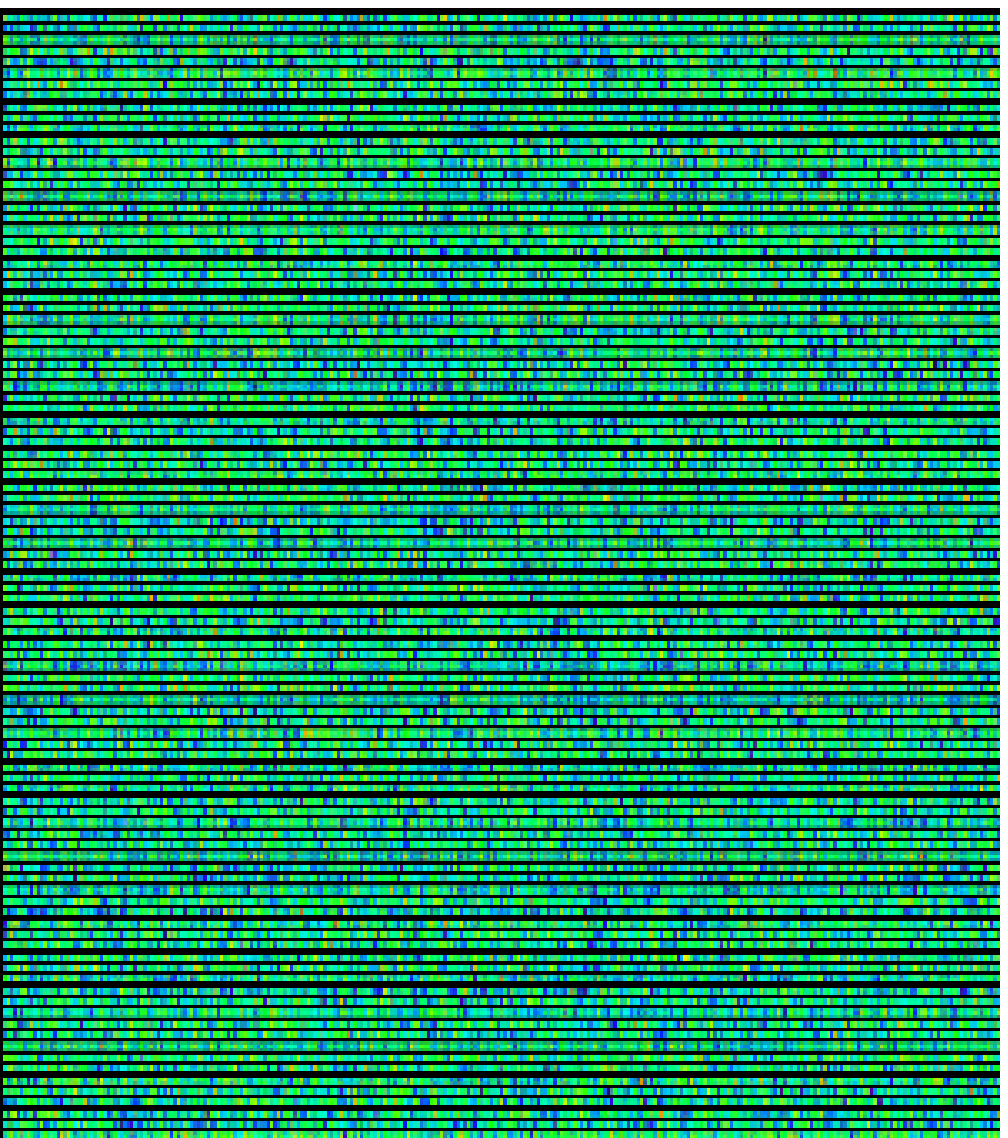}
   \includegraphics[trim=0 0 0 0,height=7.45CM,angle=0]{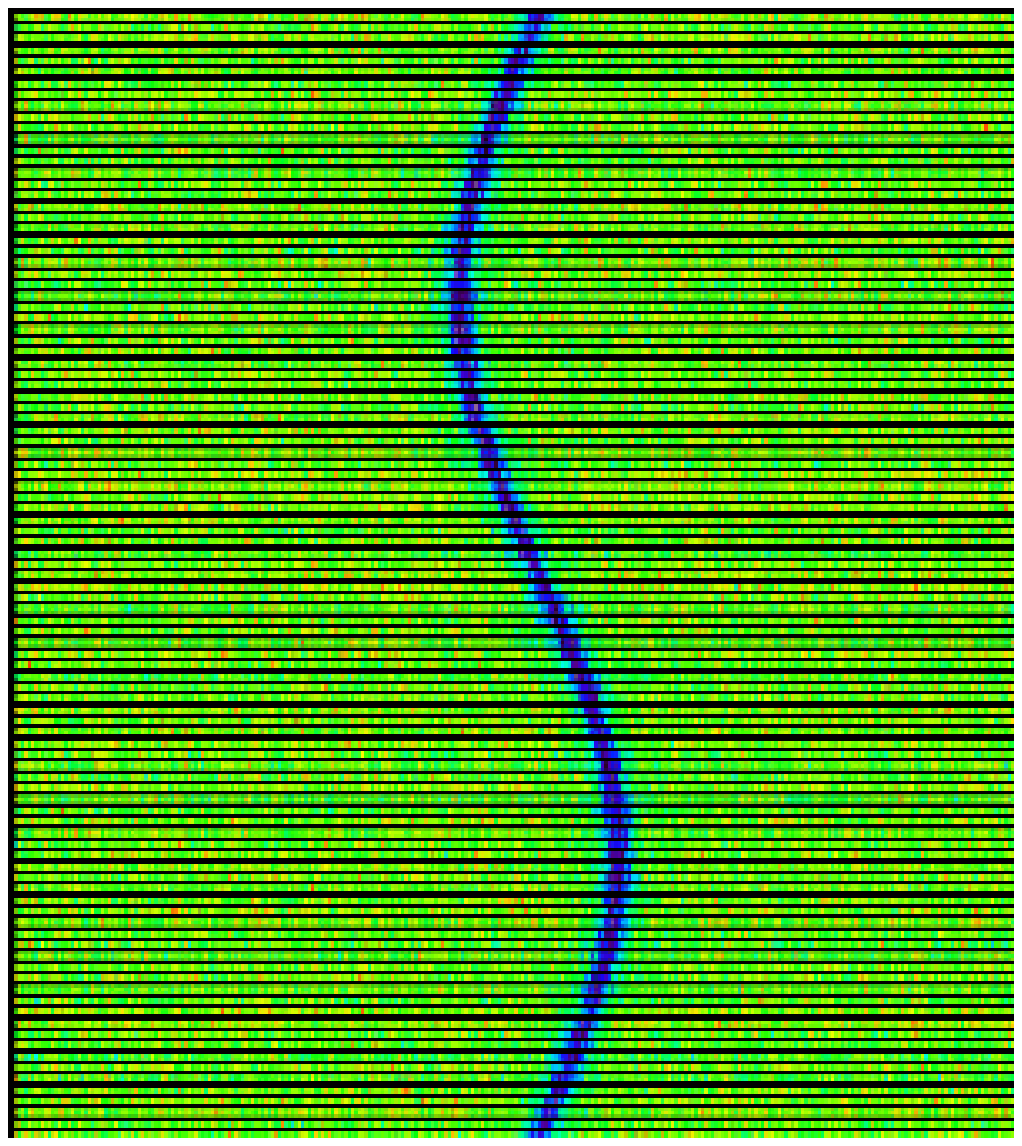}
   \includegraphics[trim=0 0 0 0,height=7.45CM,angle=0]{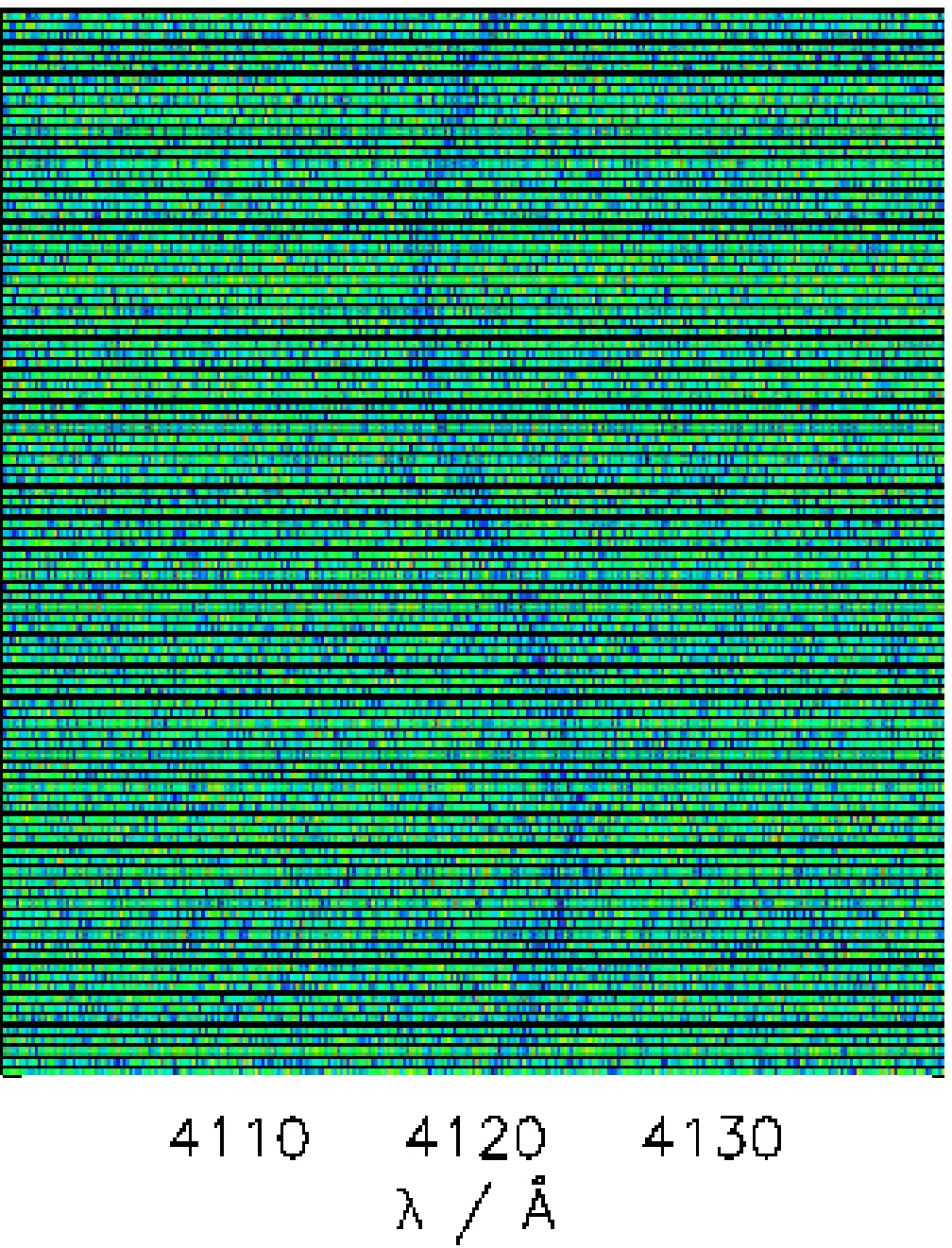}
   \includegraphics[trim=0 0 0 0,height=7.45CM,angle=0]{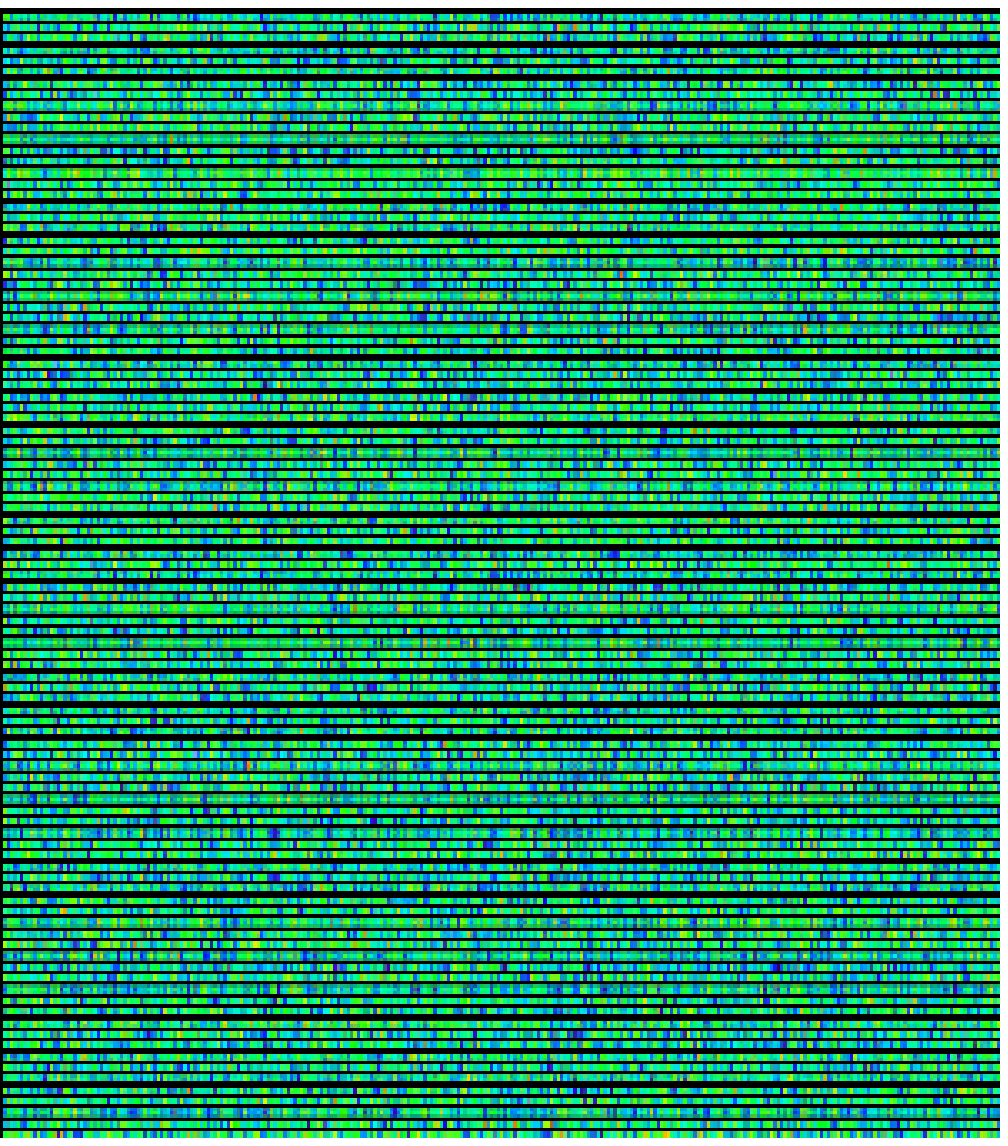}
     \caption{Detection limit for s-curves. A synthetic line profile (phase-dependently shifted with
              $K_\mathrm{sec} = 230\,\mathrm{km/s}$) at three strengths
              (equivalent widths of $W_\lambda = 122, 15, \mathrm{and}\, 3\,\mathrm{m\AA}$, from left to right)
              is shown for S/N = 50, 25 (approximately our UVES data value), and 10 (top to bottom). 
              All panels show the same wavelength and phase intervals.}
   \label{fig:detlimit}
\end{figure*}
}

TLISA is a java\footnote{\url{https://java.com/en}} tool that can be retrieved freely in its desktop version 
via \url{http://astro.uni-tuebingen.de/~TLISA}. Our XSHOOTER UVB spectra are provided there as a test data package.

\subsection{Search for spectral signatures}
\label{sect:search}

\begin{figure}
   \includegraphics[trim=0 0 0 0,width=\columnwidth,angle=0]{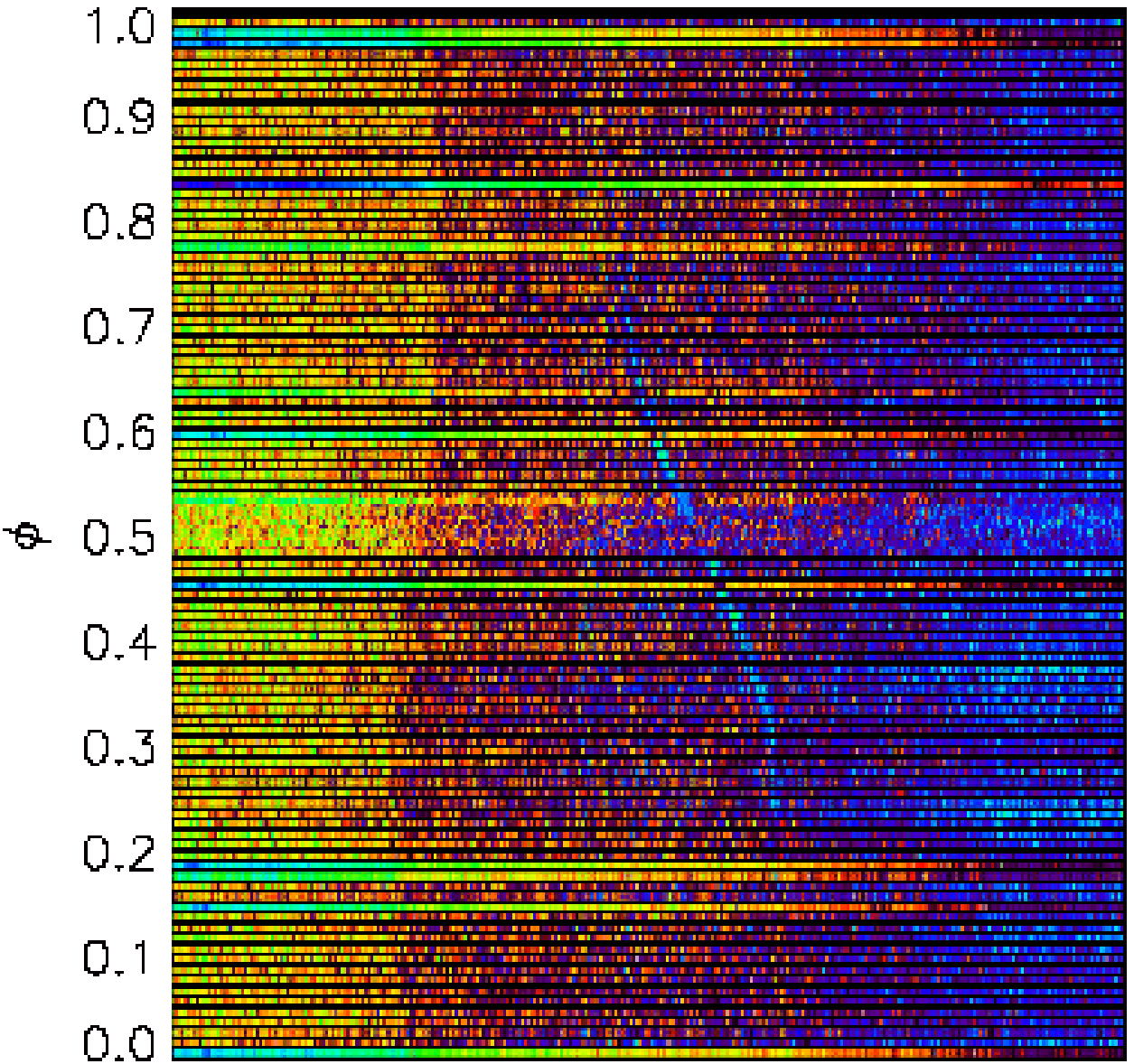}
   \includegraphics[trim=-18 0 0 0,width=\columnwidth,angle=0]{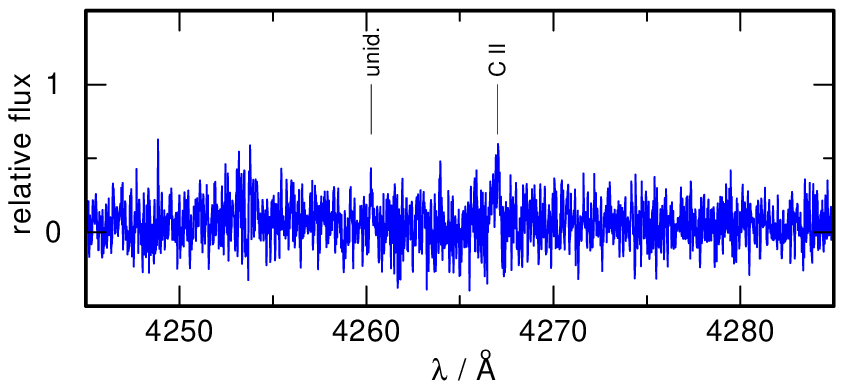}
    \caption{Top: section of the secondary's phase-dependent UVES spectra ($4245\,\mathrm{\AA} \le \lambda \le 4285\,\mathrm{\AA}$). 
             Bottom: single spectrum (2001-01-08T00-53-54.965 UT) in the secondary's rest frame in the respective wavelength interval;
             ``unid.'' denotes unidentified lines.
            }
   \label{fig:uvesa}
\end{figure}

\begin{figure}
   \includegraphics[trim=0 0 0 0,width=\columnwidth,angle=0]{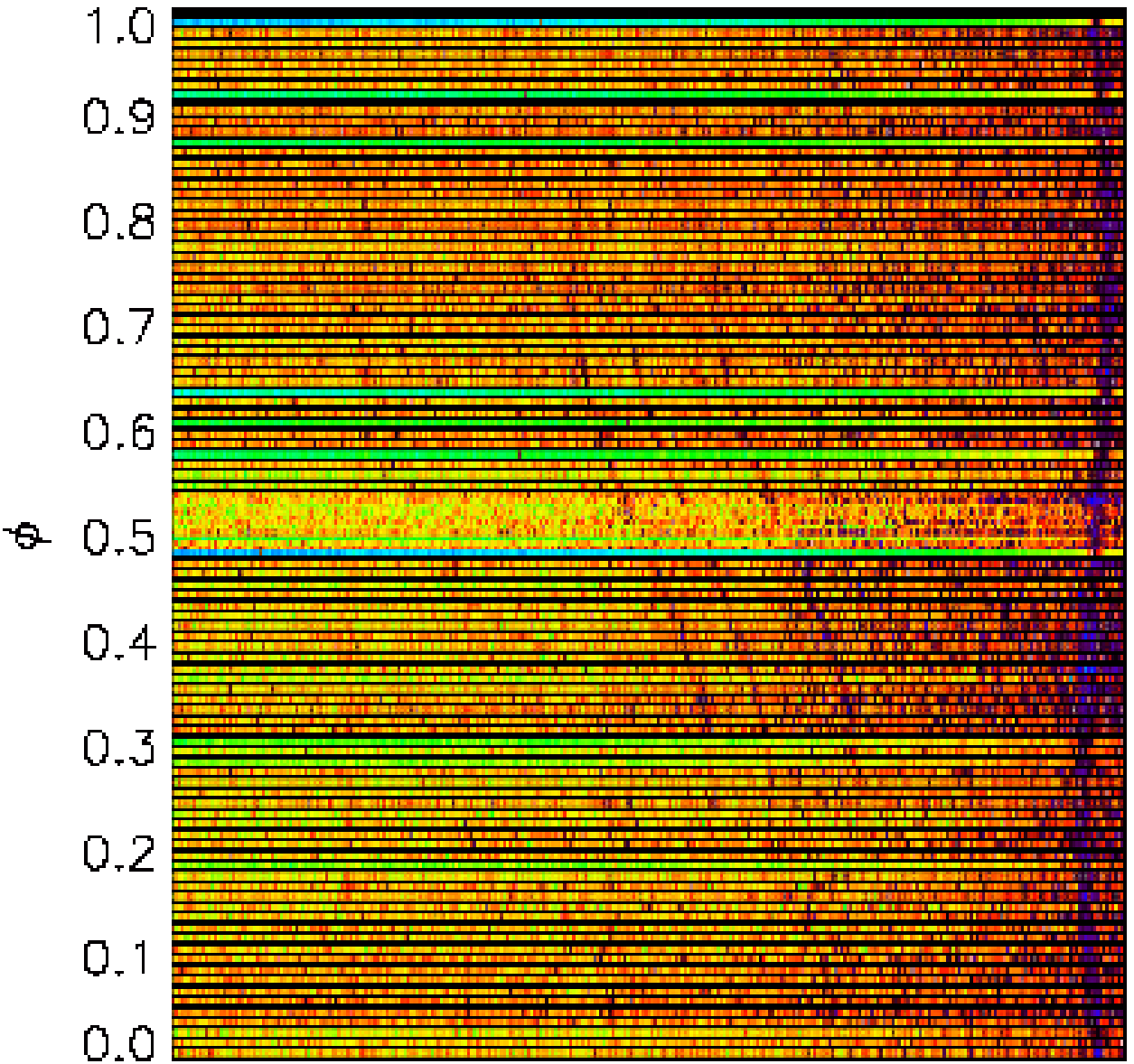}
   \includegraphics[trim=-18 0 0 0,width=\columnwidth,angle=0]{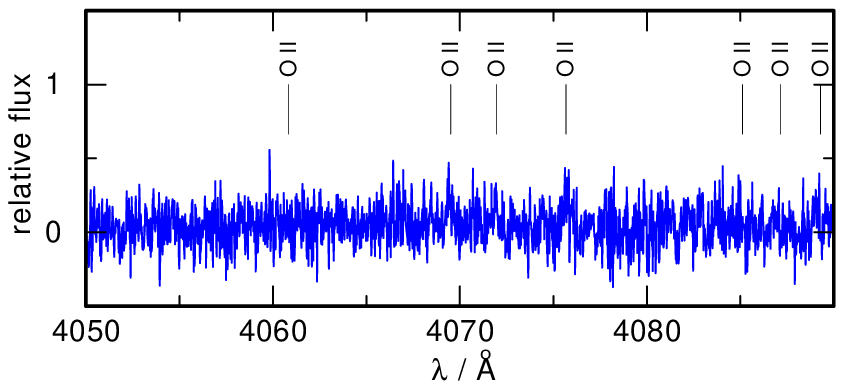}
    \caption{Same as Fig.\,\ref{fig:uvesa} for $4050\,\mathrm{\AA} \le \lambda \le 4090\,\mathrm{\AA}$. 
            }
   \label{fig:uvesb}
\end{figure}

We applied TLISA to our UVES and XSHOOTER data. 
Two representative examples for a successful search of a relatively strong and comparatively weak lines
in the UVES data are shown in Figs.\,\ref{fig:uvesa} and \ref{fig:uvesb}, respectively.
In total, we identified 73 spectral lines that stem from the secondary. 
For their unambiguous assignment, we calculated TMAP models (\Teffw{20000}, \loggw{5.0}) , which included H and one metal (C, N, or O) at solar abundance \citep{asplundetal2009}, with atomic data from the
T\"ubingen Model-Atom database (TMAD\footnote{\url{http://astro.uni-tuebingen.de/~TMAD}}).
Although the parameters do not represent the secondary, they are well suited for identification
purposes. In particular, we  modeled the multiplet line patterns precisely.
Table\,\ref{tab:lineids} shows the identified lines. We indicate lines that we did not  include in our models
due to limited model atoms, but are found in the 
National Institute of Standards and Technology
\citep{NIST_ASD}
atomic spectra database.

\begin{table}\centering
\caption{Radial-velocity amplitudes for selected lines of the secondary.}
\label{tab:ksec}
\begin{tabular}{r@{.}lr@{.}ll}
\hline\hline
\noalign{\smallskip}
\multicolumn{2}{c}{$\lambda_\mathrm{lab}\,/\,\mathrm{\AA}$} & \multicolumn{2}{c}{$K_\mathrm{sec}\,/\,\mathrm{km/s}$} & Instrument \\
\hline
3920&68 & \hbox{}\hspace{4mm}228&7 & UVES         \\
4069&62 &                    213&5 & UVES         \\
4072&15 &                    213&3 & UVES         \\
4075&86 &                    220&0 & UVES         \\
4089&29 &                    233&0 & UVES         \\
4101&74 &                    239&1 & UVES         \\
4119&22 &                    217&7 & UVES         \\
4185&44 &                    227&6 & UVES         \\
4189&79 &                    214&0 & UVES         \\
4267&09 &                    223&3 & UVES         \\
4414&91 &                    228&5 & UVES         \\
4416&97 &                    228&4 & UVES         \\
4699&22 &                    214&7 & UVES         \\
4705&35 &                    208&4 & UVES         \\
\hline                       
4069&62 &                    220&3 & XSHOOTER UVB \\
4101&74 &                    225&4 & XSHOOTER UVB \\
4267&09 &                    229&8 & XSHOOTER UVB \\
4340&47 &                    219&5 & XSHOOTER UVB \\
4414&91 &                    222&1 & XSHOOTER UVB \\
4416&97 &                    215&7 & XSHOOTER UVB \\
4861&33 &                    213&3 & XSHOOTER UVB \\
\hline                       
7065&19 &                    210&2 & XSHOOTER VIS \\
\hline
\end{tabular}
\tablefoot{
Wavelengths correspond to Table\,\ref{tab:lineids}.
}
\end{table}

\begin{figure}
   \includegraphics[trim=0 0 0 0,width=\columnwidth,angle=0]{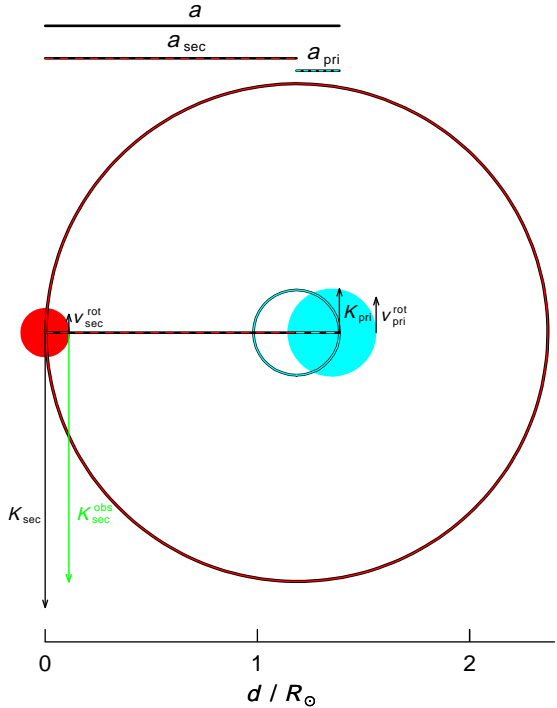}
    \caption{Dimensions (drawn to scale), orbits, and velocities of \aador. 
            }
   \label{fig:ksec}
\end{figure}

It is worthwhile to note that neither \citet{rauchwerner2003} nor 
\citet[][and priv\@. comm.]{rucinski2009} succeeded in a search for lines of the secondary.
\citet{vuckovicetal2008} identified more than 20 narrow emission lines,
mainly of \ion{C}{ii} and \ion{O}{ii} and broad Balmer emission lines with core absorption 
(Table\,\ref{tab:lineids} indicates those explicitly mentioned by the authors). 
All these groups had the same UVES data at their disposal.

We used TLISA to determine the radial-velocity amplitude of the secondary and
arrived at an average of $K^\mathrm{obs}_\mathrm{sec} = 220 \pm 10\,\mathrm{km/s}$. 
To control the quality of this graphical method, we used 
the image reduction and analysis facility (IRAF\footnote{\url{http://iraf.noao.edu},
IRAF is distributed by the National Optical Astronomy Observatory,
which is operated by the Associated Universities for Research in
Astronomy, Inc., under cooperative agreement with the National Science
Foundation.})
to measure the wavelengths shifts for selected lines in all spectra (Table\,\ref{tab:ksec}).
We measured instrument-average values of
$K^\mathrm{obs}_\mathrm{sec} = 222.2 \pm 2.5\,\mathrm{km/s}$ and
$K^\mathrm{obs}_\mathrm{sec} = 220.9 \pm 2.3\,\mathrm{km/s}$ (1\,$\sigma$ errors) from the UVES and XSHOOTER UVB data.
In the XSHOOTER VIS data, only one line could be used from which we determined
$K^\mathrm{obs}_\mathrm{sec} = 210.2 \pm 6.4\,\mathrm{km/s}$. Weighted by the number of individual
results and the 13 times better spectral resolution of UVES, we calculate an average value of
$K^\mathrm{obs}_\mathrm{sec} = 222.1 \pm 3.5\,\mathrm{km/s}$. This value is almost equal to
the TLISA value (above) and demonstrates well that TLISA is a quite powerful tool.

\begin{figure*}
   \resizebox{\hsize}{!}{\includegraphics{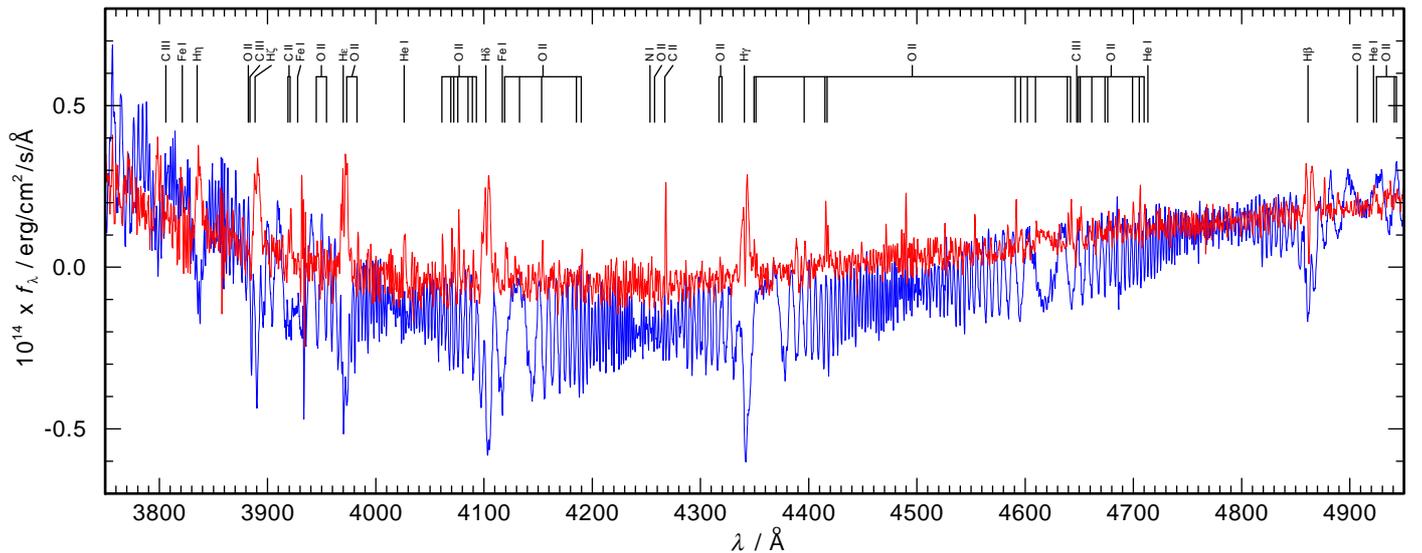}}
    \caption{Extracted XSHOOTER UVB spectra of the secondary with the primary master
             spectrum (Sect.\,\ref{sect:primary}) subtracted. 
             The spectra are processed with a Savitzky-Golay low-pass filter \citep{savitzkygolay1964}.
             Red: five spectra coadded just before occultation;
             blue: five spectra coadded just before transit.
             The identified lines (Table\,\ref{tab:lineids}) are marked.
            }
   \label{fig:specsec}
\end{figure*}

\begin{figure*}
   \resizebox{\hsize}{!}{\includegraphics{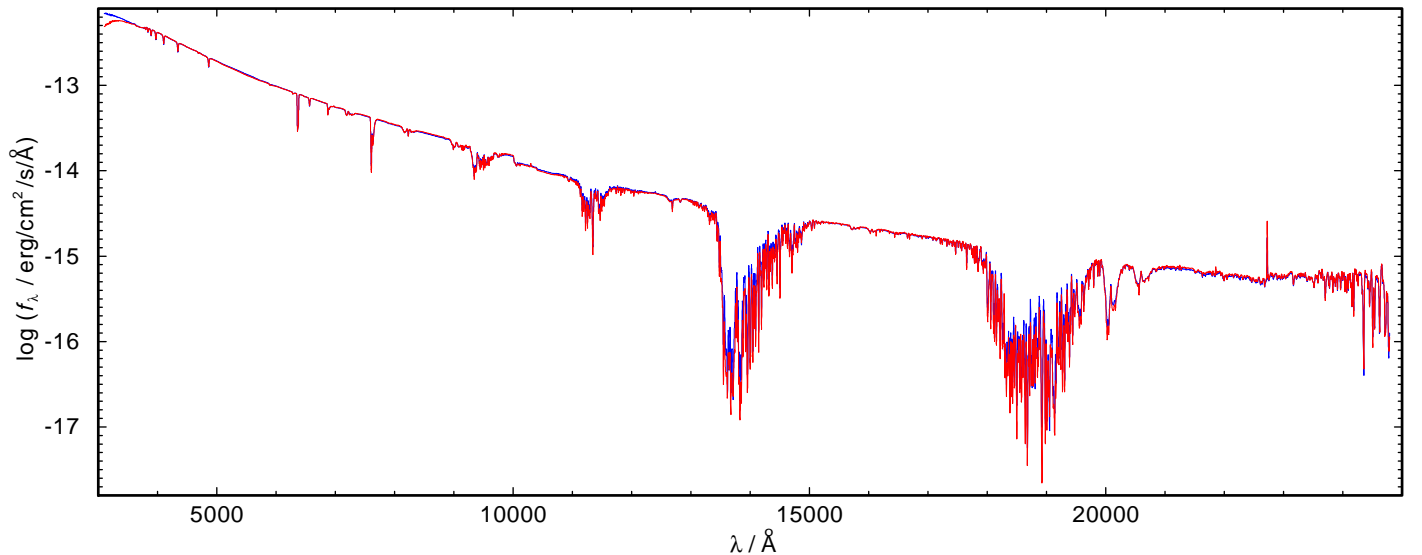}}
    \caption{XSHOOTER spectra (smoothed with a Gaussian with a FWHM of 10\,\AA\ for clarity) 
             during transit of the secondary (red, ten individual spectra next to $T_\mathrm{0}$ coadded) and 
             during its occultation (blue, three spectra coadded).
            }
   \label{fig:tr_vs_oc}
\end{figure*}

Figure\,\ref{fig:ksec} illustrates the dimensions and velocities of \aador.
All rotations and orbital motions are counterclockwise.
The parameter $K^\mathrm{obs}_\mathrm{sec}$ is measured. 

Our spectra show clearly that a day and a night side exist (Fig.~\ref{fig:scurve}). Neither
an efficient horizontal energy flow nor a significantly unsynchronized rotation of the 
secondary\footnote{The rotation of the primary is not synchronized yet \citep{muelleretal2010}.} 
yield an isothermal surface structure. 

The assumption of a synchronized rotation of the secondary and of a strict separation 
of an irradiated day side and a nonirradiated night side is a simplification \citep[cf\@.,][]{barmanetal2004}. 
In reality, horizontal energy flows (heat conduction or winds)  couples the two sides. Improved 1.5 dimensional 
(1.5\,D) modeling of irradiated planets \citep{barmanetal2005} have shown that the temperature and 
mean-intensity distribution at a planet's surface resemble a horseshoe shape \citep{hauschildtetal2008}. 
Better spectra with much higher S/N and resolution would allow us to investigate  this effect and crucially 
constrain future (3\,D) modeling. 

However, $a_\mathrm{sec}$ calculated from $K^\mathrm{obs}_\mathrm{sec}$ would be too small. 
Since no detailed model for the surface structure or phase-dependent spectra of the
secondary of \aador exist, we estimate that the center of gravity of the measured emission
lines is dominated by a region that lies toward the irradiating primary star, but not 
very much off the secondary's orbital path.
The maximum error is $K_\mathrm{sec}^\mathrm{corr} = 2\pi R_\mathrm{sec} / P$ 
(emission region lies closest to the primary). 
We assume a quarter of that value 
with respective error range 
($5.4^{+16.2}_{-5.4}\,\mathrm{km/s}$,
with
$R_\mathrm{sec} = 0.1112\,R_\odot$,
Sect.\,\ref{sect:masses}).
In addition, the secondary is assumed to perform a synchronous
rotation with $v_\mathrm{sec}^\mathrm{rot} = 2\pi R_\mathrm{sec} / P $. 
Projected at our assumed main emission region, we use
$v_\mathrm{sec}^\mathrm{rot,em} = \sin \alpha\, v_\mathrm{sec}^\mathrm{rot}$ (with $\alpha = 15\degr$).
Thus, the real orbital velocity of the secondary is

\begin{equation*}
K_\mathrm{sec} = K^\mathrm{obs}_\mathrm{sec} + v^\mathrm{rot,em}_\mathrm{sec} + K^\mathrm{corr}_\mathrm{sec}\,\,.
\end{equation*}

\noindent
In the case of \aador, we calculate 
$K_\mathrm{sec} = (222.1\pm 3.5) + (5.4\pm 0.6) + (5.4^{+16.2}_{-5.4})\,\mathrm{km/s} = 232.9^{+16.6}_{-6.5}\,\mathrm{km/s}$.
This is, within its error limits, well in agreement with the previously determined value of
$K_\mathrm{sec} \ge 230 \pm 10\,\mathrm{km/s}$ given by \citet{vuckovicetal2008}
and the estimate of $K_\mathrm{sec} = 240 \pm 20\,\mathrm{km/s}$ by \citet{muelleretal2010}.

\subsection{Spectrum of the secondary}
\label{sect:secspec}

We successfully extracted the spectrum of the secondary's irradiated side
(Fig.\,\ref{fig:specsec})
from those XSHOOTER UVB spectra where \ionw{C}{ii}{4267.09} (Fig.\,\ref{fig:scurve})
is prominent. Although we observed spectrophotometric standard stars at the beginning 
and the end of the night, the observing conditions that were widely varying on short
timescales during our XSHOOTER night 
(see \url{http://astro.uni-tuebingen.de/~rauch/ESO_092.C-0692.gif} for an impression)
hampered the flux calibration and, thus, our ability to achieve the predicted, better quality even 
in the coadded spectrum. 

In the XSHOOTER UVB, VIS, and NIR spectra (Fig.\,\ref{fig:tr_vs_oc}), we did not find any  significant, unambiguous spectral
features of a (nonirradiated) low-mass dwarf star \citep[e.g.,][]{lietal2014,littlefairetal2014,manjavakasetal2014}. In this phase, the expected secondary's contribution to the composite spectrum of \aador is of the order of 
0.1\,\% in the NIR. Our UVES and XSHOOTER spectra are not good enough to measure this.
The small deviations between the individual observations stem most likely from variations of the telluric 
absorption during the night.

\subsection{Inspection of the FUSE observations}
\label{sect:fuse}

\begin{figure}
   \includegraphics[trim=0 0 0 0,width=\columnwidth,angle=0]{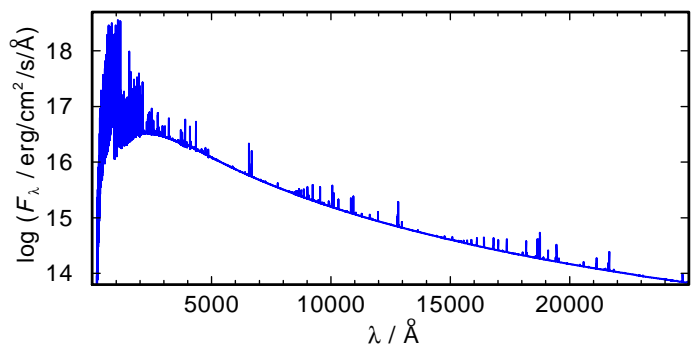}
    \caption{Synthetic spectrum of the irradiated hemisphere of the secondary of \aador.
            }
   \label{fig:fuse}
\end{figure}

The PHOENIX\footnote{\url{http://www.hs.uni-hamburg.de/EN/For/ThA/phoenix}} 
\citep{hauschildtbaron2009,hauschildtbaron2010,phoenix2010}
model-atmosphere calculations 
for the irradiated secondary by \citet{barmanetal2004} show that its
strongest lines are located in the ultraviolet
(Fig.\,\ref{fig:fuse}), especially in the FUSE wavelength range ($910\,\mathrm{\AA} \la \lambda \la 1188\mathrm{\AA}$).

\begin{figure}
   \includegraphics[trim=0 0 0 0,width=\columnwidth,angle=0]{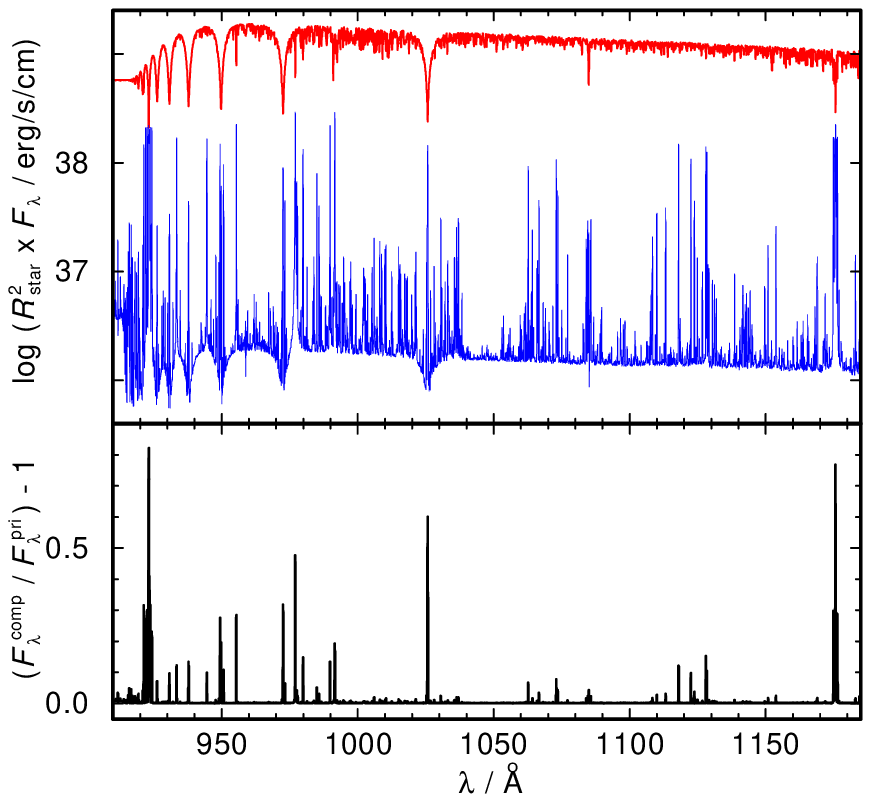}
    \caption{Top panel:
             contributions of the primary and the secondary to the composite flux of \aador 
             in the FUSE wavelength range,
             convolved with rotational profiles using $v^\mathrm{rot}_\mathrm{pri} = 31.8$\,km/s 
             \citep{rucinski2009} and $v^\mathrm{rot}_\mathrm{sec} = 21.5$\,km/s (calculated from 
             $R_\mathrm{sec}=0.1112$, Sect.\,\ref{sect:masses}).  The FUSE resolution was simulated
             by a subsequent convolution with a Gaussian (FWHM = 0.05\,\AA). 
             Bottom panel: ratio of the composite flux to the primary's flux. 
            }
   \label{fig:fuse_comp}
\end{figure}

To calculate the composite spectrum of \aador, the synthetic spectra
of the primary \citep{klepprauch2011} and
the secondary \citep{barmanetal2004} were scaled with 
$R^2_\mathrm{pri}$ and 
$R^2_\mathrm{sec}$
(Sect.\,\ref{sect:masses}), respectively. 
The contribution of the secondary to the composite spectrum is shown in Fig.\,\ref{fig:fuse_comp}.

\begin{figure}\centering
   \resizebox{0.5\hsize}{!}{\includegraphics{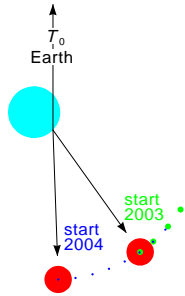}}
    \caption{Same as Fig.\,\ref{fig:obstimes}, for our FUSE observations
             (2003: four big, green dots,
              2008: eight small, blue dots).
            }
   \label{fig:obstimes_fuse}
\end{figure}

Therefore, we analyzed the 12 individual spectra (planned exposure time of 200\,s each) that we obtained with FUSE
(ObsId D0250101000, 2003-08-29, four observations, total exposure time of 731\,s,
       D0250102000, 2004-06-22, eight observations, 1623\,s).
They cover about an eighth of the orbit of \aador (Fig.\,\ref{fig:obstimes_fuse}).

The procedure for analyzing the FUSE data was similar to that described above for the UVES and XSHOOTER spectra. 
First, we measured the velocity offsets from one exposure to the next from the ISM features and coaligned the 
exposures on these. Then, we measured velocity offsets of the photospheric features 
(\ion{C}{iii}, \ion{Si}{iv}, and \ion{P}{v}) 
relative to the ISM lines for each exposure and shifted the exposures in velocity space to coalign on the 
photospheric features.The phase coverage, while not complete,  adequately  determined the orbital phase 
for each exposure and confirms the phase predicted from the measured period.  The phase-aligned spectra were 
then combined to produce a template of the primary star. As described above, any features of the secondary 
star spectrum should be widely dispersed in velocity space and should leave no imprint on this template. 
This template was  shifted to match the velocity of the primary star in each exposure and then subtracted. 
We computed the velocity offset for the secondary star relative to the primary, based on our orbital solution 
(Sect.\,\ref{sect:period}) and orbital velocity (Sect.\,\ref{sect:search}).  We then shifted each 
template-subtracted spectrum in velocity space by the predicted velocity offset of the secondary star, and 
combined the resulting spectra.  
Unfortunately, no features obviously appear to increase in significance in the coadded FUSE spectrum. 
Even an individual search for \ion{C}{ii-iii} and \ion{O}{ii-iii} lines remains entirely negative, 
although the spectra were taken in both years directly after the secondary’s occultation (Fig.\,\ref{fig:obstimes_fuse}) 
where the emission lines are strongest.
This nondetection is unexplained. It appears possible that the contribution of the secondary to the 
composite spectrum of \aador in the FUSE wavelength range is overestimated by the current model 
(Fig.\,\ref{fig:fuse_comp}).
A more thorough exploration of these models is beyond the scope of the present paper and 
will be the subject of a future study.

\section{Orbit dimensions, masses, and radii}
\label{sect:masses}

With both radial-velocity amplitudes known, 
$K_\mathrm{pri} = 40.15 \pm 0.11\,\mathrm{km/s}$ \citep{muelleretal2010} and
$K_\mathrm{sec} = 232.9^{+16.6}_{-6.5},\mathrm{km/s}$,
we determined the dimensions, masses, and radii of the components of \aador.
Their orbital radii and the distance of the stars were calculated with

\begin{eqnarray*}
\label{eq:axis}
        a_\mathrm{pri} & = \frac{P \cdot K_\mathrm{pri}}{2\pi} & = 0.2074 \pm 0.0006\,R_\odot \\
        a_\mathrm{sec} & = \frac{P \cdot K_\mathrm{sec}}{2\pi} & = 1.2029^{+0.0857}_{-0.0336}\,R_\odot \\ 
        a             & = a_\mathrm{pri} + a_\mathrm{sec}      & = 1.4102^{+0.0863}_{-0.0341}\,R_\odot\,\,.
\end{eqnarray*}

\noindent
The total mass of the system is

\begin{equation*}
\label{eq:sysmass}
        M = \frac{4\pi^2a^3}{G\cdot P^2} = 0.5516^{+0.1076}_{-0.0341}\,M_\odot
\end{equation*}

\noindent
(with the gravitational constant $G$). With $K_\mathrm{pri} M_\mathrm{pri} = K_\mathrm{sec} M_\mathrm{sec}$, the components' masses are

\begin{eqnarray*}
\label{eq:commass}
        M_\mathrm{pri} & = 0.4705^{+0.0975}_{-0.0354}\,M_\odot\,\,\,\, \\
        M_\mathrm{sec} & = 0.0811^{+0.0184}_{-0.0102}\,M_\odot\,\,.
\end{eqnarray*}

\noindent
The radius of the primary is

\begin{equation*}
\label{eq:radpri}
    R_\mathrm{pri} = \sqrt{M_\mathrm{pri}\cdot G / g_\mathrm{pri}} = 0.2113^{+0.0346}_{-0.0195}\,R_\odot\,\, ,  
\end{equation*}

\noindent
where $g_\mathrm{pri}$ is its surface gravity.
\citet{kilkennyetal1979} determined $R_\mathrm{pri} / R_\mathrm{sec} = 1.9$. Therefore the secondary's
radius is 
$R_\mathrm{sec} = 0.1112^{+0.0182}_{-0.0102}\,R_\odot$.

\section{Results and conclusions}
\label{sect:results}

The extracted secondary's spectrum allows us to investigate  its nature. 
A cool brown dwarf 
shows, for $T_\mathrm{eff} \lesssim 1200$\,K, a typical spectrum that is dominated by 
NH$_3$ and CH$_4$ bands \citep[see, for example,][]{burninghametal2008}.
In contrast, a late M-type dwarf exhibits, e.g\@., TiO bands
\citep{kirkpatricketal1993}. We did not succeed in extracting the spectrum of the
cool side of the secondary.
However, both types exhibit emission lines of C\,{\sc ii} and O\,{\sc ii} 
\citep[as already identified by][]{vuckovicetal2008}
only due to irradiation of the primary and thus, the heating of one hemisphere.

We have identified 73 lines of the secondary (Table\,\ref{tab:lineids}), i.e., the number of identified lines was 
increased by about a factor of three. This was possible because of the consequent implementation
of the s-curve method in the VO tool TLISA, which  allows us to search even for very weak lines. 
The identified lines stem mainly from \ion{C}{ii - iii} and \ion{O}{ii}. 

To make progress on the understanding of the nature
of the secondary and, hence, the whole system \aador,
further UVES observations in its red arm ($4200\,\mathrm{\AA} \ge \lambda \ge 11000\,\mathrm{\AA}$) on a dark night
with superb seeing conditions are highly desirable.

\begin{figure}
   \resizebox{\hsize}{!}{\includegraphics{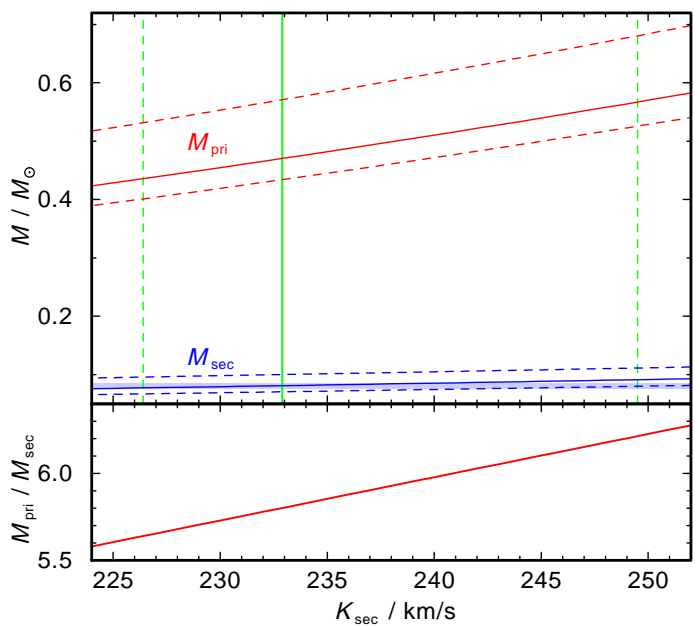}}
    \caption{Top: dependency of the components' masses (red: primary dashed lines show the error range,
             blue: secondary) on the orbital velocity of the secondary 
             \citep[for $K_\mathrm{pri} = 40.15\,\mathrm{km/s}$,][]{muelleretal2010}. 
             The light blue horizontal region indicates the hydrogen-burning mass limit \citep{chabrierbaraffe1997}.
             The green thick line shows our determination of $K_\mathrm{sec} = 232.9\,\mathrm{km/s}$.
             The error range is indicated by the green dashed lines.
            }
   \label{fig:massratio}
\end{figure}

We improved the period measurement of \citet{rauchwerner2003} because of the extended time interval
and verified the period of $P = 22597.03320768 \pm 0.0000691\,\mathrm{s}$, which was determined from
the light curve of \aador by \citet{kilkenny2011}.
We measured an orbital velocity of 
$K_\mathrm{sec} = 232.9^{+16.6}_{-6.5}\,\mathrm{km/s}$
for the secondary.
We determined masses
of 
$M_\mathrm{pri} = 0.4705^{+0.0975}_{-0.0354}\,M_\odot$
for the primary star and of 
$M_\mathrm{sec} = 0.0811^{+0.0184}_{-0.0102}\,M_\odot$
for the secondary star.
The secondary's mass is fully overlapping (Fig.\,\ref{fig:massratio}) 
with the hydrogen-burning mass limit \citep[$0.075 - 0.085$\,\Msol,][]{chabrierbaraffe1997} and, thus, it is 
still not possible to decide about the nature of the object.

\begin{figure}
   \resizebox{\hsize}{!}{\includegraphics{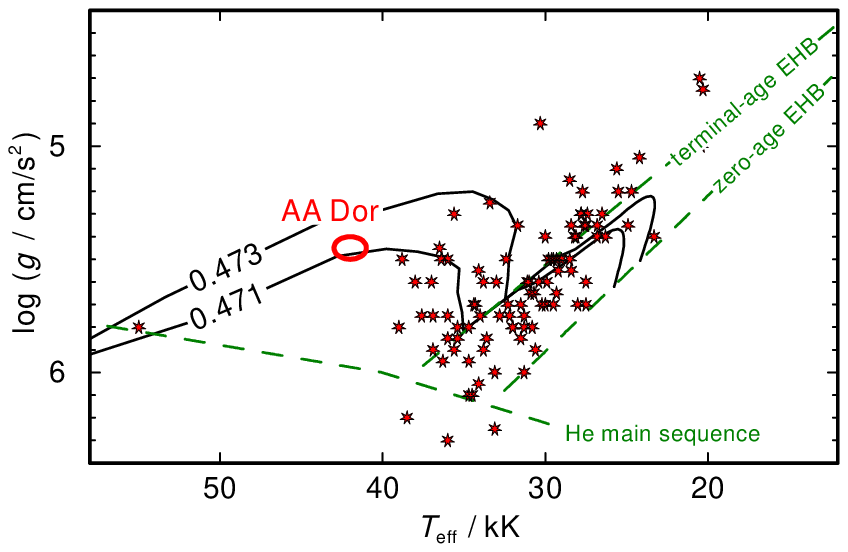}}
   \caption{Location of the primary of \aador in the \Teff $-$ \logg
            plane \citep[the ellipse indicates the error limits given by][]{klepprauch2011}
            compared to sdBs and sdOBs from \citet{edelmann2003}.
            Post-EHB tracks from \citet[full black lines, labeled with the
            respective stellar masses in $M_\odot$, solar H/He ratio and solar metallicity on the
            HB]{dormanetal1993} are also shown for comparison.
        }
\label{fig:evol}
\end{figure}

Our primary and secondary masses agree perfectly with those of 
\citet[$M_\mathrm{pri} = 0.4714 \pm 0.0050\,M_\odot$, $M_\mathrm{sec} = 0.0788^{+0.0075}_{-0.0063}\,M_\odot$]{klepprauch2011}.
They assumed a solar He abundance ($Y = 0.288$) on the horizontal branch (HB) and determined $M_\mathrm{pri}$ by a comparison 
with respective post-EHB evolutionary tracks (Fig.\,\ref{fig:evol}) of \citet{dormanetal1993}.
Since \aador had experienced a common-envelope event and its present abundance pattern (Table\,\ref{tab:abund}) exhibits
the interplay of gravitational settling and radiative levitation \citep[cf\@.][]{rauch2000},
any information about its former photospheric abundances is lost. However, the assumption of solar abundances
for \aador on the HB appears reasonable.

Our masses also agree well  with those of
       \citet[][$0.33\,M_\odot  \le M_\mathrm{pri} \le 0.47\,M_\odot$ and 
                $0.064\,M_\odot \le M_\mathrm{sec} \le 0.082\,M_\odot$]{hilditchetal2003}.
The values of \citet[$M_\mathrm{pri} = 0.510^{+0.125}_{-0.108}\,M_\odot$, 
                     $M_\mathrm{sec} = 0.085^{+0.031}_{-0.023}\,M_\odot$]{muelleretal2010}
agree with ours within error limits.

Detailed 3D modeling of the phase-dependent spectrum of the irradiated secondary of \aador that considers 
horizontal energy flows is highly desirable to verify our assumptions for the correction of the observed 
$K^\mathrm{obs}_\mathrm{sec}$ in Sect.\,\ref{sect:search}. Our primary mass 
$M_\mathrm{pri} = 0.47\,M_\odot$, which was calculated with $K_\mathrm{sec} = 232.9,\mathrm{km/s}$, agrees 
with the mass interval of $0.33\,M_\odot  \le M_\mathrm{pri} \le 0.47\,M_\odot$ 
\citep[][from light-curve analysis]{hilditchetal2003}. Higher values of $K_\mathrm{sec}$ yield higher values of $M_\mathrm{pri}$
(Fig.\,\ref{fig:massratio}). E.g., $K_\mathrm{sec} = 240,\mathrm{km/s}$ results in
$M_\mathrm{pri} = 0.5104^{+0.1061}_{-0.0424}\,M_\odot$ which is, even within error limits, outside the 
mass interval (see above).

\begin{acknowledgements}
DH and TR are supported by the German Aerospace Center (DLR, grants 50\,OR\,1501 and 05\,OR\,1402, respectively).
The GAVO project at T\"ubingen was supported by the Federal Ministry of Education and
Research (BMBF, grants  05\,AC\,6\,VTB, 05\,AC\,11\,VTB).
The TLISA tool (\url{http://astro.uni-tuebingen.de/~TLISA}) used for this paper was constructed as part 
of the activities of the German Astrophysical Virtual Observatory.
The UVES and XSHOOTER spectra used in this analysis were obtained as parts of ESO Service Mode runs,
programs 066.D-1800 and 092.C-0692, respectively. 
We thank 
David Kilkenny for his help in determining the exact time of the secondary's occultation during our XSHOOTER observations,
Roger Wesson who had performed these observations and successfully covered this critical time, and
Travis Barman who put the irradiated spectrum of the secondary at our disposal.
Some of the data presented in this paper were obtained from the
Mikulski Archive for Space Telescopes (MAST). STScI is operated by the
Association of Universities for Research in Astronomy, Inc., under NASA
contract NAS5-26555. Support for MAST for non-HST data is provided by
the NASA Office of Space Science via grant NNX09AF08G and by other
grants and contracts. 
This research has made use of NASA's Astrophysics Data System and
the SIMBAD database, operated at CDS, Strasbourg, France.
\end{acknowledgements}

\bibliographystyle{aa}
\bibliography{26229}

\begin{thebibliography}{62}
\expandafter\ifx\csname natexlab\endcsname\relax\def\natexlab#1{#1}\fi

\bibitem[{{Almeida} {et~al.}(2012){Almeida}, {Jablonski}, {Tello}, \&
  {Rodrigues}}]{almeidaetal2012}
{Almeida}, L.~A., {Jablonski}, F., {Tello}, J., \& {Rodrigues}, C.~V. 2012,
  \mnras, 423, 478

\bibitem[{{Asplund} {et~al.}(2009){Asplund}, {Grevesse}, {Sauval}, \&
  {Scott}}]{asplundetal2009}
{Asplund}, M., {Grevesse}, N., {Sauval}, A.~J., \& {Scott}, P. 2009, \araa, 47,
  481

\bibitem[{{Barlow} {et~al.}(2013){Barlow}, {Kilkenny}, {Drechsel}, {Dunlap},
  {O'Donoghue}, {Geier}, {O'Steen}, {Clemens}, {LaCluyze}, {Reichart},
  {Haislip}, {Nysewander}, \& {Ivarsen}}]{barlowetal2013}
{Barlow}, B.~N., {Kilkenny}, D., {Drechsel}, H., {et~al.} 2013, \mnras, 430, 22

\bibitem[{{Barman} {et~al.}(2004){Barman}, {Hauschildt}, \&
  {Allard}}]{barmanetal2004}
{Barman}, T.~S., {Hauschildt}, P.~H., \& {Allard}, F. 2004, \apj, 614, 338

\bibitem[{{Barman} {et~al.}(2005){Barman}, {Hauschildt}, \&
  {Allard}}]{barmanetal2005}
{Barman}, T.~S., {Hauschildt}, P.~H., \& {Allard}, F. 2005, \apj, 632, 1132

\bibitem[{{Baron} {et~al.}(2010){Baron}, {Chen}, \& {Hauschildt}}]{phoenix2010}
{Baron}, E., {Chen}, B., \& {Hauschildt}, P.~H. 2010, {PHOENIX: A
  General-purpose State-of-the-art Stellar and Planetary Atmosphere Code},
  Astrophysics Source Code Library

\bibitem[{{Bergeron} {et~al.}(1993){Bergeron}, {Wesemael}, {Lamontagne}, \&
  {Chayer}}]{bergeronetal1993}
{Bergeron}, P., {Wesemael}, F., {Lamontagne}, R., \& {Chayer}, P. 1993, \apjl,
  407, L85

\bibitem[{{Burningham} {et~al.}(2008){Burningham}, {Pinfield}, {Leggett},
  {Tamura}, {Lucas}, {Homeier}, {Day-Jones}, {Jones}, {Clarke}, {Ishii},
  {Kuzuhara}, {Lodieu}, {Zapatero Osorio}, {Venemans}, {Mortlock}, {Barrado Y
  Navascu{\'e}s}, {Martin}, \& {Magazz{\`u}}}]{burninghametal2008}
{Burningham}, B., {Pinfield}, D.~J., {Leggett}, S.~K., {et~al.} 2008, \mnras,
  391, 320

\bibitem[{{Chabrier} \& {Baraffe}(1997)}]{chabrierbaraffe1997}
{Chabrier}, G. \& {Baraffe}, I. 1997, \aap, 327, 1039

\bibitem[{{Chabrier} {et~al.}(2000){Chabrier}, {Baraffe}, {Allard}, \&
  {Hauschildt}}]{charbrieretal2000}
{Chabrier}, G., {Baraffe}, I., {Allard}, F., \& {Hauschildt}, P. 2000, \apj,
  542, 464

\bibitem[{{Dorman} {et~al.}(1993){Dorman}, {Rood}, \&
  {O'Connell}}]{dormanetal1993}
{Dorman}, B., {Rood}, R.~T., \& {O'Connell}, R.~W. 1993, \apj, 419, 596

\bibitem[{{Edelmann}(2003)}]{edelmann2003}
{Edelmann}, H. 2003, Dissertation, {University of Erlangen-Nuremberg}

\bibitem[{{Fitzpatrick}(1999)}]{fitzpatrick1999}
{Fitzpatrick}, E.~L. 1999, \pasp, 111, 63

\bibitem[{{Fleig} {et~al.}(2008){Fleig}, {Rauch}, {Werner}, \&
  {Kruk}}]{fleigetal2008}
{Fleig}, J., {Rauch}, T., {Werner}, K., \& {Kruk}, J.~W. 2008, \aap, 492, 565

\bibitem[{{Geckeler} {et~al.}(2014){Geckeler}, {Schuh}, {Dreizler}, {Deetjen},
  {Gleissner}, {Risse}, {Rauch}, {G{\"o}hler}, {H{\"u}gelmeyer}, {Husser},
  {Israel}, {Benlloch-Garcia}, {Pottschmidt}, \& {Wilms}}]{geckeleretal2014}
{Geckeler}, R.~D., {Schuh}, S., {Dreizler}, S., {et~al.} 2014, {TRIPP: Time
  Resolved Imaging Photometry Package}, Astrophysics Source Code Library

\bibitem[{{Geier} {et~al.}(2010){Geier}, {Heber}, {Podsiadlowski}, {Edelmann},
  {Napiwotzki}, {Kupfer}, \& {M{\"u}ller}}]{geieretal2010}
{Geier}, S., {Heber}, U., {Podsiadlowski}, P., {et~al.} 2010, \aap, 519, A25

\bibitem[{{Hauschildt} {et~al.}(2008){Hauschildt}, {Barman}, \&
  {Baron}}]{hauschildtetal2008}
{Hauschildt}, P.~H., {Barman}, T., \& {Baron}, E. 2008, Physica Scripta Volume
  T, 130, 014033

\bibitem[{{Hauschildt} \& {Baron}(2009)}]{hauschildtbaron2009}
{Hauschildt}, P.~H. \& {Baron}, E. 2009, \aap, 498, 981

\bibitem[{{Hauschildt} \& {Baron}(2010)}]{hauschildtbaron2010}
{Hauschildt}, P.~H. \& {Baron}, E. 2010, \aap, 509, A36

\bibitem[{{Heber}(2009)}]{heber2009}
{Heber}, U. 2009, \araa, 47, 211

\bibitem[{{Hilditch} {et~al.}(1996){Hilditch}, {Harries}, \&
  {Hill}}]{hilditchetal1996}
{Hilditch}, R.~W., {Harries}, T.~J., \& {Hill}, G. 1996, \mnras, 279, 1380

\bibitem[{{Hilditch} {et~al.}(2003){Hilditch}, {Kilkenny}, {Lynas-Gray}, \&
  {Hill}}]{hilditchetal2003}
{Hilditch}, R.~W., {Kilkenny}, D., {Lynas-Gray}, A.~E., \& {Hill}, G. 2003,
  \mnras, 344, 644

\bibitem[{{Kilkenny}(2011)}]{kilkenny2011}
{Kilkenny}, D. 2011, \mnras, 412, 487

\bibitem[{{Kilkenny} {et~al.}(1978){Kilkenny}, {Hilditch}, \&
  {Penfold}}]{kilkennyetal1978}
{Kilkenny}, D., {Hilditch}, R.~W., \& {Penfold}, J.~E. 1978, \mnras, 183, 523

\bibitem[{{Kilkenny} {et~al.}(1979){Kilkenny}, {Hilditch}, \&
  {Penfold}}]{kilkennyetal1979}
{Kilkenny}, D., {Hilditch}, R.~W., \& {Penfold}, J.~E. 1979, \mnras, 187, 1

\bibitem[{{Kirkpatrick} {et~al.}(1993){Kirkpatrick}, {Kelly}, {Rieke},
  {Liebert}, {Allard}, \& {Wehrse}}]{kirkpatricketal1993}
{Kirkpatrick}, J.~D., {Kelly}, D.~M., {Rieke}, G.~H., {et~al.} 1993, \apj, 402,
  643

\bibitem[{{Klepp} \& {Rauch}(2011)}]{klepprauch2011}
{Klepp}, S. \& {Rauch}, T. 2011, \aap, 531, L7

\bibitem[{Kramida {et~al.}(2014)Kramida, {Yu.~Ralchenko}, Reader, \& {and NIST
  ASD Team}}]{NIST_ASD}
Kramida, A., {Yu.~Ralchenko}, Reader, J., \& {and NIST ASD Team}. 2014, {NIST
  Atomic Spectra Database (ver. 5.2), [Online]. Available:
  {\url{http://physics.nist.gov/asd}} [\today]. National Institute of Standards
  and Technology, Gaithersburg, MD.}

\bibitem[{{Li} {et~al.}(2014){Li}, {Zhang}, {Han}, {Kong}, \&
  {Gong}}]{lietal2014}
{Li}, L., {Zhang}, F., {Han}, Q., {Kong}, X., \& {Gong}, X. 2014, \mnras, 445,
  1331

\bibitem[{{Littlefair} {et~al.}(2014){Littlefair}, {Casewell}, {Parsons},
  {Dhillon}, {Marsh}, {G{\"a}nsicke}, {Bloemen}, {Catalan}, {Irawati}, {Hardy},
  {Mcallister}, {Bours}, {Richichi}, {Burleigh}, {Burningham}, {Breedt}, \&
  {Kerry}}]{littlefairetal2014}
{Littlefair}, S.~P., {Casewell}, S.~L., {Parsons}, S.~G., {et~al.} 2014,
  \mnras, 445, 2106

\bibitem[{{Livio} \& {Soker}(1984)}]{liviosoker1984}
{Livio}, M. \& {Soker}, N. 1984, \mnras, 208, 783

\bibitem[{{Manjavacas} {et~al.}(2014){Manjavacas}, {Bonnefoy}, {Schlieder},
  {Allard}, {Rojo}, {Goldman}, {Chauvin}, {Homeier}, {Lodieu}, \&
  {Henning}}]{manjavakasetal2014}
{Manjavacas}, E., {Bonnefoy}, M., {Schlieder}, J.~E., {et~al.} 2014, \aap, 564,
  A55

\bibitem[{{Maxted} {et~al.}(2000){Maxted}, {North}, \&
  {Marsh}}]{maxtedetal2000}
{Maxted}, P.~F.~L., {North}, R.~C., \& {Marsh}, T.~R. 2000, The Newsletter of
  the Isaac Newton Group of Telescopes, 3, 7

\bibitem[{{M{\"u}ller} {et~al.}(2010){M{\"u}ller}, {Geier}, \&
  {Heber}}]{muelleretal2010}
{M{\"u}ller}, S., {Geier}, S., \& {Heber}, U. 2010, \apss, 329, 101

\bibitem[{{Napiwotzki} \& {Rauch}(1994)}]{napiwotzkirauch1994}
{Napiwotzki}, R. \& {Rauch}, T. 1994, \aap, 285, 603

\bibitem[{{Rauch}(2000)}]{rauch2000}
{Rauch}, T. 2000, \aap, 356, 665

\bibitem[{{Rauch}(2004)}]{rauch2004}
{Rauch}, T. 2004, \apss, 291, 275

\bibitem[{{Rauch}(2012)}]{rauch2012a}
{Rauch}, T. 2012, in Astronomical Society of the Pacific Conference Series,
  Vol. 452, Fifth Meeting on Hot Subdwarf Stars and Related Objects, ed.
  D.~{Kilkenny}, C.~S. {Jeffery}, \& C.~{Koen}, 111

\bibitem[{{Rauch} \& {Deetjen}(2003)}]{rauchdeetjen2003}
{Rauch}, T. \& {Deetjen}, J.~L. 2003, in Astronomical Society of the Pacific
  Conference Series, Vol. 288, Stellar Atmosphere Modeling, ed. I.~{Hubeny},
  D.~{Mihalas}, \& K.~{Werner}, 103

\bibitem[{{Rauch} \& {Werner}(2003)}]{rauchwerner2003}
{Rauch}, T. \& {Werner}, K. 2003, \aap, 400, 271

\bibitem[{{Rauch} {et~al.}(2012){Rauch}, {Werner}, {Bi{\'e}mont}, {Quinet}, \&
  {Kruk}}]{rauchetal2012ge}
{Rauch}, T., {Werner}, K., {Bi{\'e}mont}, {\'E}., {Quinet}, P., \& {Kruk},
  J.~W. 2012, \aap, 546, A55

\bibitem[{{Rauch} {et~al.}(2014{\natexlab{a}}){Rauch}, {Werner}, {Quinet}, \&
  {Kruk}}]{rauchetal2014zn}
{Rauch}, T., {Werner}, K., {Quinet}, P., \& {Kruk}, J.~W. 2014{\natexlab{a}},
  \aap, 564, A41

\bibitem[{{Rauch} {et~al.}(2014{\natexlab{b}}){Rauch}, {Werner}, {Quinet}, \&
  {Kruk}}]{rauchetal2014ba}
{Rauch}, T., {Werner}, K., {Quinet}, P., \& {Kruk}, J.~W. 2014{\natexlab{b}},
  \aap, 566, A10

\bibitem[{{Rauch} {et~al.}(2015){Rauch}, {Werner}, {Quinet}, \&
  {Kruk}}]{rauchetal2015ga}
{Rauch}, T., {Werner}, K., {Quinet}, P., \& {Kruk}, J.~W. 2015, \aap, 577, A6

\bibitem[{{Ritter}(1986)}]{ritter1986}
{Ritter}, H. 1986, \aap, 169, 139

\bibitem[{{Ritter} \& {Kolb}(2003)}]{ritterkolb2003}
{Ritter}, H. \& {Kolb}, U. 2003, \aap, 404, 301

\bibitem[{{Rucinski}(2009)}]{rucinski2009}
{Rucinski}, S.~M. 2009, \mnras, 395, 2299

\bibitem[{{Savitzky} \& {Golay}(1964)}]{savitzkygolay1964}
{Savitzky}, A. \& {Golay}, M.~J.~E. 1964, Analytical Chemistry, 36, 1627

\bibitem[{{Schaffenroth} {et~al.}(2015){Schaffenroth}, {Barlow}, {Drechsel}, \&
  {Dunlap}}]{schaffenrothetal2015}
{Schaffenroth}, V., {Barlow}, B.~N., {Drechsel}, H., \& {Dunlap}, B.~H. 2015,
  ArXiv 1502.04459

\bibitem[{{Schaffenroth} {et~al.}(2013){Schaffenroth}, {Geier}, {Drechsel},
  {Heber}, {Wils}, {{\O}stensen}, {Maxted}, \& {di
  Scala}}]{schaffenrothetal2013}
{Schaffenroth}, V., {Geier}, S., {Drechsel}, H., {et~al.} 2013, \aap, 553, A18

\bibitem[{{Schaffenroth} {et~al.}(2014){Schaffenroth}, {Geier}, {Heber},
  {Kupfer}, {Ziegerer}, {Heuser}, {Classen}, \&
  {Cordes}}]{schaffenrothetal2014}
{Schaffenroth}, V., {Geier}, S., {Heber}, U., {et~al.} 2014, \aap, 564, A98

\bibitem[{{Schreiber} \& {G{\"a}nsicke}(2003)}]{schreibergaensicke2003}
{Schreiber}, M.~R. \& {G{\"a}nsicke}, B.~T. 2003, \aap, 406, 305

\bibitem[{{Schuh} {et~al.}(2003){Schuh}, {Dreizler}, {Deetjen}, \&
  {G{\"o}hler}}]{schuhetal2003}
{Schuh}, S.~L., {Dreizler}, S., {Deetjen}, J.~L., \& {G{\"o}hler}, E. 2003,
  Baltic Astronomy, 12, 167

\bibitem[{{Scott} {et~al.}(2015{\natexlab{a}}){Scott}, {Asplund}, {Grevesse},
  {Bergemann}, \& {Sauval}}]{scottetal2015b}
{Scott}, P., {Asplund}, M., {Grevesse}, N., {Bergemann}, M., \& {Sauval}, A.~J.
  2015{\natexlab{a}}, \aap, 573, A26

\bibitem[{{Scott} {et~al.}(2015{\natexlab{b}}){Scott}, {Grevesse}, {Asplund},
  {Sauval}, {Lind}, {Takeda}, {Collet}, {Trampedach}, \&
  {Hayek}}]{scottetal2015a}
{Scott}, P., {Grevesse}, N., {Asplund}, M., {et~al.} 2015{\natexlab{b}}, \aap,
  573, A25

\bibitem[{{Tremblay} \& {Bergeron}(2009)}]{tremblaybergeron2009}
{Tremblay}, P.-E. \& {Bergeron}, P. 2009, \apj, 696, 1755

\bibitem[{{Vernet} {et~al.}(2011){Vernet}, {Dekker}, {D'Odorico}, {Kaper},
  {Kjaergaard}, {Hammer}, {Randich}, {Zerbi}, {Groot}, {Hjorth}, {Guinouard},
  {Navarro}, {Adolfse}, {Albers}, {Amans}, {Andersen}, {Andersen}, {Binetruy},
  {Bristow}, {Castillo}, {Chemla}, {Christensen}, {Conconi}, {Conzelmann},
  {Dam}, {de Caprio}, {de Ugarte Postigo}, {Delabre}, {di Marcantonio},
  {Downing}, {Elswijk}, {Finger}, {Fischer}, {Flores}, {Fran{\c c}ois},
  {Goldoni}, {Guglielmi}, {Haigron}, {Hanenburg}, {Hendriks}, {Horrobin},
  {Horville}, {Jessen}, {Kerber}, {Kern}, {Kiekebusch}, {Kleszcz}, {Klougart},
  {Kragt}, {Larsen}, {Lizon}, {Lucuix}, {Mainieri}, {Manuputy}, {Martayan},
  {Mason}, {Mazzoleni}, {Michaelsen}, {Modigliani}, {Moehler}, {M{\o}ller},
  {Norup S{\o}rensen}, {N{\o}rregaard}, {P{\'e}roux}, {Patat}, {Pena}, {Pragt},
  {Reinero}, {Rigal}, {Riva}, {Roelfsema}, {Royer}, {Sacco}, {Santin},
  {Schoenmaker}, {Spano}, {Sweers}, {Ter Horst}, {Tintori}, {Tromp}, {van
  Dael}, {van der Vliet}, {Venema}, {Vidali}, {Vinther}, {Vola}, {Winters},
  {Wistisen}, {Wulterkens}, \& {Zacchei}}]{vernetetal2011}
{Vernet}, J., {Dekker}, H., {D'Odorico}, S., {et~al.} 2011, \aap, 536, A105

\bibitem[{{Vu{\v c}kovi{\'c}} {et~al.}(2008){Vu{\v c}kovi{\'c}}, {{\O}stensen},
  {Bloemen}, {Decoster}, \& {Aerts}}]{vuckovicetal2008}
{Vu{\v c}kovi{\'c}}, M., {{\O}stensen}, R., {Bloemen}, S., {Decoster}, I., \&
  {Aerts}, C. 2008, in Astronomical Society of the Pacific Conference Series,
  Vol. 392, Hot Subdwarf Stars and Related Objects, ed. U.~{Heber}, C.~S.
  {Jeffery}, \& R.~{Napiwotzki}, 199

\bibitem[{{Werner}(1996)}]{werner1996}
{Werner}, K. 1996, \apjl, 457, L39

\bibitem[{{Werner} {et~al.}(2003){Werner}, {Deetjen}, {Dreizler}, {Nagel},
  {Rauch}, \& {Schuh}}]{werneretal2003}
{Werner}, K., {Deetjen}, J.~L., {Dreizler}, S., {et~al.} 2003, in Astronomical
  Society of the Pacific Conference Series, Vol. 288, Stellar Atmosphere
  Modeling, ed. I.~{Hubeny}, D.~{Mihalas}, \& K.~{Werner}, 31

\bibitem[{{Werner} {et~al.}(2012{\natexlab{a}}){Werner}, {Dreizler}, \&
  {Rauch}}]{tmap2012}
{Werner}, K., {Dreizler}, S., \& {Rauch}, T. 2012{\natexlab{a}}, {TMAP:
  T{\"u}bingen NLTE Model-Atmosphere Package}, Astrophysics Source Code Library

\bibitem[{{Werner} {et~al.}(2012{\natexlab{b}}){Werner}, {Rauch}, {Ringat}, \&
  {Kruk}}]{werneretal2012}
{Werner}, K., {Rauch}, T., {Ringat}, E., \& {Kruk}, J.~W. 2012{\natexlab{b}},
  \apjl, 753, L7

\end{thebibliography}

\onecolumn

\begin{longtable}{r@{.}lr@{.}llcc}
\caption{\label{tab:lineids}Identified lines of the secondary in \aador.
                            The subscripts obs and lab denote observed (at $\phi = 0.5$) and laboratory-measured wavelengths, respectively.
                            In case  lines are found in both spectra, $\lambda_\mathrm{obs}$ is measured in the UVES spectrum.
                            Columns 4 and 5 indicate in which spectrum they are identified.} \\
\hline\hline
\noalign{\smallskip}
\multicolumn{2}{c}{$\lambda_\mathrm{obs}$\,/ \,\AA} & \multicolumn{2}{c}{$\lambda_\mathrm{lab}$\,/ \,\AA} & Ion & UVES & XSHOOTER \\
\noalign{\smallskip}
\hline
\endfirsthead
\caption{continued.}\\
\hline\hline
\noalign{\smallskip}
\multicolumn{2}{c}{$\lambda_\mathrm{obs}$\,/ \,\AA} & \multicolumn{2}{c}{$\lambda_\mathrm{lab}$\,/ \,\AA} & Ion & UVES & XSHOOTER \\
\noalign{\smallskip}
\hline
\noalign{\smallskip}
\endhead
\hline
\noalign{\smallskip}
\endfoot
\noalign{\smallskip}
 3805&79 &  3806&12                  & \ion{C}{iii}\tablefootmark{c}   & $\times$ &  UVB \\
 3820&61 &  3821&18                  & \ion{Fe}{i}\tablefootmark{b,c}  & $\times$ &  UVB \\
 3835&01 &  3835&39                  & H\,$\eta$                       & $\times$ &  UVB \\
 3881&96 &  3882&19                  & \ion{O}{ii}\tablefootmark{c}    & $\times$ &      \\
 3883&83 &  3883&82                  & \ion{C}{iii}\tablefootmark{c}   & $\times$ &      \\
 3888&21 &  3889&02                  & H\,$\zeta$                      & $\times$ &  UVB \\
 3918&83 &  3918&97                  & \ion{C}{ii}                     & $\times$ &  UVB \\
 3920&67 &  3920&68                  & \ion{C}{ii}                     & $\times$ &  UVB \\
 3927&27 &  3927&74                  & \ion{Fe}{i}\tablefootmark{b,c}  &          &  UVB \\
 3944&80 &  3945&04                  & \ion{O}{ii}                     & $\times$ &      \\
 3954&14 &  3954&36                  & \ion{O}{ii}                     & $\times$ &  UVB \\
 3969&95 &  3970&08                  & H\,$\epsilon$                   & $\times$ &  UVB \\
 3973&07 &  3973&26                  & \ion{O}{ii}                     & $\times$ &      \\
 4026&32 &  4026&21                  & \ion{He}{i}                     & $\times$ &  UVB \\
 4060&84 &  4061&03                  & \ion{O}{ii}\tablefootmark{b}    & $\times$ &      \\
 4069&45 &  4069&62                  & \ion{O}{ii}                     & $\times$ &  UVB \\
 4072&07 &  4072&15                  & \ion{O}{ii}                     & $\times$ &  UVB \\
 4075&77 &  4075&86                  & \ion{O}{ii}                     & $\times$ &  UVB \\
 4085&05 &  4085&11                  & \ion{O}{ii}                     & $\times$ &  UVB \\
 4087&15 &  4087&15                  & \ion{O}{ii}                     & $\times$ &  UVB \\
 4089&05 &  4089&29                  & \ion{O}{ii}\tablefootmark{b,c}  & $\times$ &  UVB \\
 4092&75 &  4092&93                  & \ion{O}{ii}                     & $\times$ &      \\
 4101&69 &  4101&74                  & H\,$\delta$\tablefootmark{a}    & $\times$ &  UVB \\
 4116&64 &  4116&95                  & \ion{Fe}{i}\tablefootmark{b,c}  & $\times$ &      \\
 4119&02 &  4119&22                  & \ion{O}{ii}                     & $\times$ &      \\
 4132&52 &  4132&80                  & \ion{O}{ii}                     & $\times$ &      \\
 4185&30 &  4185&44                  & \ion{O}{ii}\tablefootmark{b}    & $\times$ &      \\
 4189&76 &  4189&79                  & \ion{O}{ii}\tablefootmark{b}    & $\times$ &      \\
 4253&60 &  4253&39                  & \ion{N}{i }                     & $\times$ &  UVB \\
 4257&34 &  4257&54                  & \ion{O}{ii}                     & $\times$ &  UVB \\
 4267&40 &  4267&09                  & \ion{C}{ii}\tablefootmark{a}    & $\times$ &  UVB \\
 4317&02 &  4317&14                  & \ion{O}{ii}                     & $\times$ &  UVB \\
 4319&79 &  4319&87                  & \ion{O}{ii}\tablefootmark{b}    & $\times$ &  UVB \\
 4340&56 &  4340&47                  & H\,$\gamma$                     & $\times$ &  UVB \\
 4349&41 &  4349&43                  & \ion{O}{ii}\tablefootmark{a}    & $\times$ &      \\
 4351&25 &  4351&26                  & \ion{O}{ii}                     & $\times$ &      \\
 4395&75 &  4395&93                  & \ion{O}{ii}                     &          &  UVB \\
 4415&02 &  4414&91                  & \ion{O}{ii}\tablefootmark{a}    & $\times$ &  UVB \\
 4416&95 &  4416&97                  & \ion{O}{ii}\tablefootmark{a}    & $\times$ &  UVB \\
 4590&81 &  4590&97                  & \ion{O}{ii}                     & $\times$ &  UVB \\
 4595&96 &  4595&96                  & \ion{O}{ii}                     & $\times$ &  UVB \\
 4602&43 &  4602&13                  & \ion{O}{ii}                     &          &  UVB \\
 4609&66 &  4609&44                  & \ion{O}{ii}\tablefootmark{b,c}  & $\times$ &  UVB \\
 4638&74 &  4638&86                  & \ion{O}{ii}                     & $\times$ &  UVB \\
 4641&81 &  4641&81                  & \ion{O}{ii}                     & $\times$ &  UVB \\
 4647&28 &  4647&72                  & \ion{C}{iii}\tablefootmark{b,c} & $\times$ &  UVB \\
 4649&05 &  4649&13                  & \ion{O}{ii}                     & $\times$ &  UVB \\
 4650&59 &  4650&84                  & \ion{O}{ii}                     & $\times$ &  UVB \\
 4661&67 &  4661&63                  & \ion{O}{ii}                     & $\times$ &      \\
 4673&93 &  4673&73                  & \ion{O}{ii}\tablefootmark{c}    & $\times$ &      \\
 4676&08 &  4676&24                  & \ion{O}{ii}                     & $\times$ &      \\
 4699&01 &  4699&22                  & \ion{O}{ii}                     & $\times$ &      \\
 4705&16 &  4705&35                  & \ion{O}{ii}                     & $\times$ &      \\
 4710&09 &  4710&01                  & \ion{O}{ii}\tablefootmark{b}    & $\times$ &      \\
 4713&64 &  4713&38                  &  \ion{He}{i}                    & $\times$ &      \\
 4861&38 &  4861&33                  & H\,$\beta$                      & $\times$ &  UVB \\
 4906&86 &  4906&83                  & \ion{O}{ii}\tablefootmark{b}    &          &  UVB \\
 4921&97 &  4921&93                  & \ion{He}{i}                     & $\times$ &  UVB \\
 4924&52 &  4924&53                  & \ion{O}{ii}\tablefootmark{b,c}  & $\times$ &  UVB \\
 4940&97 &  4941&07                  & \ion{O}{ii}                     & $\times$ &  UVB \\
 4942&82 &  4943&01                  & \ion{O}{ii}                     & $\times$ &  UVB \\
 5875&63 &  5875&62                  & \ion{He}{i}                     &          &  VIS \\
 6151&31 &  6151&62                  & \ion{Fe}{i}\tablefootmark{b,c}  &          &  VIS \\
 6562&82 &  6562&79                  & H\,$\alpha$                     &          &  VIS \\
 6578&06 &  6578&05                  & \ion{C}{ii}                     &          &  VIS \\
 6582&90 &  6582&88                  & \ion{C}{ii}                     &          &  VIS \\
 6780&09 &  6780&50                  & \ion{C}{ii}                     &          &  VIS \\
 6783&83 &  6783&91                  & \ion{C}{ii}                     &          &  VIS \\
 6787&38 &  6787&21                  & \ion{C}{ii}                     &          &  VIS \\
 7065&32 &  7065&19                  & \ion{He}{i}                     &          &  VIS \\
 8558&13 &  8558&52                  & \ion{Fe}{i}\tablefootmark{b,c}  &          &  VIS \\
 9903&74 &  9904&06                  & \ion{Fe}{ii}\tablefootmark{b,c} &          &  VIS \\
17654&46 & 17654&53                  & \ion{O}{ii}                     &          &  NIR \\
\hline
\multicolumn{7}{p{8.5cm}}{}\vspace{-9mm} \\
\multicolumn{7}{p{8.5cm}}{
\tablefoot{
\tablefoottext{a}{identified and mentioned by \citet{vuckovicetal2008},}
\tablefoottext{b}{NIST wavelength, line not included in our models,}         
\tablefoottext{c}{uncertain}           
}
} \\
\noalign{\smallskip}
\end{longtable}

\twocolumn

\end{document}